\documentclass[a4paper,onecolumn,unpublished]{quantumarticle}
\pdfoutput=1
\usepackage[utf8]{inputenc}
\usepackage[english]{babel}
\usepackage[T1]{fontenc}
\usepackage{float}
\usepackage[10pt]{moresize}
\usepackage{amsmath}
\usepackage{mathtools}
\usepackage{hyperref}
\usepackage{cleveref}

\usepackage{array}
\usepackage{multirow,colortbl}
\newcommand{\PreserveBackslash}[1]{\let\temp=\\#1\let\\=\temp}
\newcolumntype{C}[1]{>{\PreserveBackslash\centering}p{#1}}
\newcolumntype{R}[1]{>{\PreserveBackslash\raggedleft}p{#1}}
\newcolumntype{L}[1]{>{\PreserveBackslash\raggedright}p{#1}}
\makeatletter
\newcommand{\thickhline}{%
	\noalign {\ifnum 0=`}\fi \hrule height 1pt
	\futurelet \reserved@a \@xhline
}
\newcolumntype{"}{@{\hskip\tabcolsep\vrule width 1pt\hskip\tabcolsep}}
\makeatother

\usepackage{tikz}
\usetikzlibrary{decorations,snakes,decorations.markings,positioning,shadows.blur,calc}

\pgfmathdeclarefunction{gauss}{2}{%
	\pgfmathparse{1/(#2*sqrt(2*pi))*exp(-((\x-#1)^2)/(2*#2^2))}%
}
\pgfmathdeclarefunction{newcos}{0}{%
	\pgfmathparse{1/2*(cos(1000*(\x-1.75))+1)}%
}

\newcommand{\ket}[1]{\lvert #1 \rangle}

\definecolor{LightGray}{RGB}{220,220,220}
\definecolor{LinkColor}{RGB}{167,20,49}
\definecolor{myred}{RGB}{236, 17, 0}
\definecolor{myblue}{RGB}{10, 88, 153}
\definecolor{myorange}{RGB}{236, 137, 0}
\definecolor{mygreen}{RGB}{26, 152, 81}

\crefname{defn}{Definition}{Definitions}
\Crefname{defn}{Definition}{Definitions}
\crefname{thm}{Theorem}{Theorems}
\Crefname{thm}{Theorem}{Theorems}
\crefname{claim}{Claim}{Claims}
\Crefname{claim}{Claim}{Claims}
\crefname{lem}{Lemma}{Lemmas}
\Crefname{lem}{Lemma}{Lemmas}
\crefname{rem}{Remark}{Remarks}
\Crefname{rem}{Remark}{Remarks}
\crefname{prop}{Proposition}{Propositions}
\Crefname{prop}{Proposition}{Propositions}
\crefname{cor}{Corollary}{Corollaries}
\Crefname{cor}{Corollary}{Corollaries}
\crefname{section}{Section}{Sections}
\Crefname{section}{Section}{Sections}
\crefname{equation}{}{}
\Crefname{equation}{}{}
\crefname{figure}{Figure}{Figures}
\Crefname{figure}{Figure}{Figures}
\crefname{appendix}{Appendix}{Appendices}
\Crefname{appendix}{Appendix}{Appendices}
\crefname{table}{Table}{Tables}
\Crefname{table}{Table}{Tables}
\crefname{exmp}{Example}{Examples}
\Crefname{exmp}{Example}{Examples}
\crefname{footnote}{Footnote}{Footnote}
\Crefname{footnote}{Footnote}{Footnote}

\hypersetup{
	colorlinks,
	linkcolor=LinkColor,
	citecolor=LinkColor,
	filecolor=LinkColor,
	urlcolor=LinkColor,
	pdftitle={Quantum Information Security},
	pdfauthor={Renato Renner and Ramona Wolf}
}

\begin{document}

\title{Quantum Advantage in Cryptography}

\author{Renato Renner}
\email{renner@ethz.ch}
\affiliation{Institute for Theoretical Physics, ETH Zürich, 8093 Zürich, Switzerland}
\author{Ramona Wolf}
\email{rawolf@ethz.ch}
\affiliation{Institute for Theoretical Physics, ETH Zürich, 8093 Zürich, Switzerland}

\maketitle

\fontfamily{lmr}\selectfont

\begin{abstract}
Ever since its inception, cryptography has been caught in a vicious circle: Cryptographers keep inventing methods to hide information, and cryptanalysts break them, prompting cryptographers to invent even more sophisticated encryption schemes, and so on. But could it be that quantum information technology breaks this circle? At first sight, it looks as if it just lifts the competition between cryptographers and cryptanalysts to the next level. Indeed, quantum computers will render most of today's public key cryptosystems insecure. Nonetheless, there are good reasons to believe that cryptographers will ultimately prevail over cryptanalysts. Quantum cryptography allows us to build communication schemes whose secrecy relies only on the laws of physics and some minimum assumptions about the cryptographic hardware---leaving basically no room for an attack. While we are not yet there, this article provides an overview of the principles and state of the art of quantum cryptography, as well as an assessment of current challenges and prospects for overcoming them. 
\end{abstract}


\section{Introduction}

The art of encryption is as old as the concept of written information. For thousands of years, humans have been using more and more elaborate schemes to hide the content of messages for a variety of purposes, for example, to facilitate secret communication between governments and militaries. But those who wanted to get hold of the secrets did not remain passive, of course: They have come up with increasingly refined methods to get access to encrypted information. Consequently, history is full of examples illustrating how the most sophisticated encryption schemes were rendered useless by the ingenuity of brilliant code-breakers \cite{Kahn1996}. Is it inevitable that history will always repeat itself? Is it impossible to develop an encryption scheme that cannot be broken? 

It is undoubtedly true that currently employed encryption standards such as the RSA cryptosystem \cite{Rivest1978} and the Diffie-Hellman key exchange method \cite{Diffie1976,Merkle1978} can, in principle, be broken. They are only \emph{computationally secure}, which means that breaking them requires the ability to solve a computationally hard problem. In the case of RSA, for instance, the underlying problem is factoring large numbers into prime factors. For large enough numbers, the expected amount of computational power required to factor them within reasonable time lies beyond the capabilities of state-of-the-art computers. However, as technology  advances, what is deemed ``large enough'' must be constantly adjusted accordingly to ensure the security of RSA encryption. Hence, secrets that are encrypted with the RSA cryptosystem today might be decodable with tomorrow's computers!

Alongside developments in the field of classical computers, the imminent advent of quantum computers poses an additional threat to the security of current encryption standards. Quantum computers exploit the unique features of quantum mechanics and, as such, allow for new kinds of algorithms. Consequently, tasks that are believed to be hard for classical computers can become feasible if one has access to a universal quantum computer. Most famously, quantum computers can factor large numbers efficiently using Shor's algorithm \cite{Shor1994} and, therefore, once they are realized, render current cryptographic standards such as RSA completely insecure. To counter the potential threat that quantum computers pose on currently used cryptographic schemes, the field of post-quantum cryptography aims at developing classical encryption schemes that are secure even against quantum computers. They are also only computationally secure, but rely on problems that are believed to be hard for classical \emph{and} quantum computers. However, these will not break the vicious circle mentioned before as an example from the NIST standardization process for post-quantum algorithms \cite{NIST} demonstrates: One of the alternative finalists, called SIKE, has recently been broken within one hour on a single core classical computer \cite{SIKE2022}. This example demonstrates the need for cryptographic schemes that are provably unbreakable since, otherwise, we still run the risk that someone discovers an algorithm to crack them.

Fortunately, we do not have to fear that all ciphers will eventually be broken, not even with the help of a quantum computer. It is well-understood how to design an encryption scheme between two parties that remains secure, even if the adversary has all the (classical and quantum) computational power in the universe. We call such a cryptographic scheme \emph{information-theoretically secure}. This term encompasses the fact that this kind of security can be expressed in terms of purely information-theoretic concepts, in contrast to computational security (which requires the notion of computational complexity). The crucial ingredient is a \emph{cryptographic key} or \emph{secret key}, which is a sequence of shared random bits only known to the two parties that want to exchange private messages. If the key is uniformly random, kept perfectly secret to everyone except the two parties, and no part of it is ever reused, it can be employed in an encryption scheme called one-time pad (OTP) \cite{Vernam1926}, which is an example of an information-theoretically secure encryption protocol. To use the one-time pad, the sender adds the cryptographic key to the message they intend to encrypt. The resulting ciphertext is sent to the receiver, who subtracts the key from the ciphertext, thus retrieving the original message. Even though an adversary has access to the ciphertext and knows the general encryption method (namely the one-time pad), they cannot learn anything about the message: If the key fulfills the properties mentioned above, the ciphertext is independent of the message, thus it does not reveal any information about it.

As a result, the difficulty in creating a secure encryption method now lies in developing a protocol that generates a cryptographic key with the desired properties. This goal cannot be achieved with purely classical algorithms since we have no private communication channel; this is where quantum theory comes into play. The unique features of quantum mechanics lend themselves extraordinarily well to the task of establishing a secret key between two parties. Firstly, the intrinsic randomness of quantum states and measurements can be exploited to generate truly random bits. Secondly, the phenomenon of entanglement allows the generation of the same set of random bits in two distant locations. Thirdly, quantum states have the property that they cannot be perfectly copied, in contrast to classical bits. If an adversary attempts to access information encoded in quantum states, they have to somehow interact with the quantum state, which in turn will lead to detectable changes in the state. Consequently, the communicating parties will be alerted to the attack before confidential information has been exchanged. The establishment of a secret key via quantum mechanics is known as \emph{quantum key distribution} (QKD) \cite{Bennett1984,Ekert1991}. The first goal of this article is to explain how the features of quantum mechanics guarantee the security of QKD.

So, is quantum cryptography the solution? The answer is: yes, to some extent. Quantum cryptography offers solutions for cryptographic tasks concerning private communication, where symmetric encryption schemes are employed. As explained above, these can be realized in an information-theoretically secure fashion by combining quantum key distribution protocols with the one-time pad. However, other cryptographic tasks, such as digital signatures, require public key encryption (for example, RSA); here, quantum cryptography cannot provide any advantageous quantum protocols. The second goal of this article is to explain the role that quantum cryptography plays in the larger context of cyber security and discuss possible applications and limitations.

Finally, although quantum cryptography is a promising research area that has increasingly attracted the industry's attention in recent years, it is still under development. Even though experimental implementations of QKD protocols are reaching ever further distances and higher key rates, the state-of-the-art achievements are still several orders of magnitude away from what is required for practical applications. Therefore, the third goal of this article is to present the landscape of QKD protocols, the progress of their respective experimental realizations and the challenges that still need to be overcome to make QKD market-ready. However, since a detailed review of the state of the art of experimental implementations of QKD is beyond the scope of this article, we focus here on the theoretical concepts and refer the interested reader to \cite{Pirandola2020}.

The remainder of the article is structured as follows: In \cref{sec:Classicalcrypto}, we explain what level of security classical cryptosystems can guarantee and where they reach their limitations, in particular with regard to the possible threats posed by quantum computers. In \cref{sec:Quantumcrypto}, we present potential applications of quantum cryptography and explain where it can provide an improvement over classical algorithms and where it is not applicable. We also discuss the goal of post-quantum cryptography and its advantages over current state-of-the-art algorithms. In \cref{sec:Quantumtheory}, we introduce basic notions of quantum theory and explain how they contribute to the security of quantum cryptography. A reader who is interested mostly in the security aspects of QKD may skip this section. In \cref{sec:QKDprotocols}, we discuss how quantum key distribution works, how to quantify security, and current progress as well as challenges in theory and experiment. In \cref{sec:Outlook} we discuss the prospects of quantum cryptography in overcoming these challenges and address some of the common criticisms of this technology.
\section{Limits of classical cryptography}
\label{sec:Classicalcrypto}

The field of cryptography spans everything concerned with privacy, authentication and confidentiality in the presence of adversarial behavior. An important subfield of cryptography is secure communication, which nowadays mainly involves electronic data such as encrypting emails and other plain-text messages, online banking security, and communication related to the military, governments, and the financial market. Typical tasks include private communication, authentication (i.e., verifying a claim of identity) and digital signatures, to name just a few examples.

Classical cryptography provides the algorithms and protocols used to achieve these tasks on a day-to-day basis, such as public-key encryption schemes for digital signatures and key exchange mechanisms for confidentiality. Although these protocols have the advantage that they are easy to implement with state-of-the-art technology, they have some disadvantages regarding the level of security they can guarantee.

\subsection{Security based on computational complexity}

The security of classical protocols is usually based on the assumption that certain problems are hard in terms of their computational complexity. Widely used algorithms that fall into this category are the Diffie-Hellman key exchange method \cite{Diffie1976,Merkle1978}, which relies on the hardness of the discrete logarithm problem, and the RSA cryptosystem \cite{Rivest1978}, whose security depends on the practical difficulty of factoring the product of two large prime numbers. Basing the security of a cryptographic system on the difficulty of mathematical problems has its issues: Although it is widely believed to be true that factoring large number on a classical computers is hard, it has not yet been proven despite decades of exhaustive research in this area. As long as the hardness of this and other problems that cryptography is based on is only a conjecture, it is always possible that an efficient algorithm is found to solve them, making cryptographic schemes built on them effectively unsafe.


With the imminent advent of quantum computing devices comes another complication that can arise when relying on the hardness of specific tasks: There are types of computations, such as quantum computation, that cannot be sorted into classical complexity classes. Although integer factorization and the discrete logarithm problem are believed to be hard for classical computers, there exists a \emph{quantum} algorithm that can solve them in polynomial runtime, namely Shor's algorithm \cite{Shor1994}. Hence, it is insecure to use the Diffie-Hellman key exchange method and RSA encryption in light of the possibility that practical quantum computers will be built. It may sound like a problem that will only impact security in the (more or less) distant future, depending on the development of quantum computers. But, in fact, it is already a serious threat to the confidentiality of data today: An adversary can store information encrypted using the RSA algorithm today in its encrypted form, and decrypt it once they have access to a quantum computer. If the data are supposed to be secret for an extended period of time (which is typical for data related to the military, intelligence agencies, or medical records), that means that quantum computers are already a threat to the security of data today, before they even have been built.

\subsection{Quantitative measures of security}
\label{subsec:quantitative}

\begin{figure}[t]
	\centering
	\begin{tikzpicture}[scale=1.1]
		\draw[thick,->,>=stealth] (0,0) -- (9,0);
		\draw[thick,->,>=stealth] (0,0) -- (0,4);
		\node at (9.4,0) {time};
		\node at (0,4.25) {$\epsilon^{\mathrm{comp}}$};
		\node[color=black] at (-0.6,0.25) {$\epsilon^\mathrm{QKD}$};
		\draw[thick, color=myred] (-0.15,0.25) -- (0,0.25) -- (1.5,0.5) -- (1.5,0.75) -- (3,1) -- (3,1.3) -- (4,1.5) -- (4,3.5) -- (9,3.5);
		\draw[thick, color=myblue] (-0.15,0.25) -- (0,0.25) -- (1.25,0.3) -- (1.25,0.5) -- (2.5,0.75) -- (2.5,1.25) -- (3.75,1.5) -- (3.75,1.8) -- (5,2) -- (5,2.3) -- (6.5,2.5) -- (6.5,2.7) -- (7.25,2.9) -- (7.25,3.15) -- (9,3.45);
		\draw[thick,color=mygreen] (-0.15,0.25) -- (9,0.25);
		\node[color=myblue] at (8.5,3) {\footnotesize post-quantum};
		\node[color=myblue] at (8.5,2.75) {\footnotesize algorithm};
		\node[color=mygreen] at (8,0.5) {\footnotesize QKD protocol};
		\node[color=myred] at (8,4) {\footnotesize RSA};
		\node[color=myred] at (8,3.75) {\footnotesize algorithm};
		\draw[color=black,dashed] (4,-0.15) -- (4,1.5);
		\node[color=black] at (4,-0.4) {\footnotesize first universal};
		\node[color=black] at (4,-0.65) {\footnotesize quantum};
		\node[color=black] at (4,-0.9) {\footnotesize computer};
		\node[color=black] at (0.8,1.5) {\footnotesize algorithmic};
		\node[color=black] at (0.8,1.25) {\footnotesize discovery};
		\draw[->,>=stealth] (1.35,1.2) to[bend left] (1.5,0.85);
		\draw[->,>=stealth] (1.5,1.4) to[bend left] (2.4,1.3);
		\node at (2,2.3) {\footnotesize evolution of};
		\node at (2,2.05) {\footnotesize hardware};
		\draw[->,>=stealth] (2.65,2) to[bend left] (3.25,1.55);
		\draw[->,>=stealth] (2.8,2.25) to[bend left] (4.25,2);
	\end{tikzpicture}
	\caption{\label{fig:quantcost}\textbf{Security of cryptographic protocols over time.} The diagram shows schematically the development of the probability $\epsilon^\mathrm{comp}$ that an encryption scheme is broken if the adversary has all the computational power in the world, as a function of time. Classical algorithms (including post-quantum ones) become increasingly insecure over time due to evolution of hardware and algorithmic discoveries. If there exists an efficient quantum algorithm for breaking it (which is the case for RSA), the scheme will immediately become insecure once the first universal quantum computer is built. The failure probability $\epsilon^\mathrm{QKD}$ of quantum key distribution (evaluated under the usual assumptions discussed in \cref{subsec:assumptions}), on the other hand, always remains the same.}
\end{figure}
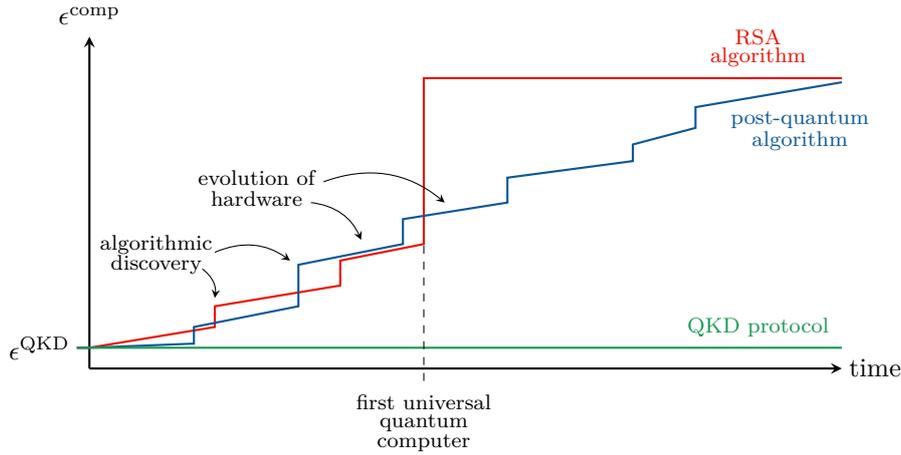

Another downside of security based on computational complexity is that it does not provide a quantitative security measure. One cannot directly map the length of the key to the level of security it provides. The decision of how long a key has to be is based on estimates of how long it takes state-of-the-art computers to crack it and how much resources they require for it. This is depicted in \cref{fig:quantcost}: The probability that a classical cryptographic scheme can be broken, denoted $\epsilon^\mathrm{comp}$, depends on its computational security. A classical algorithm, for instance the RSA encryption scheme, gradually becomes more and more insecure over time due to evolution of hardware and discoveries regarding algorithms. Maintaining the same level of security therefore becomes more and more costly, for example because one has to use longer keys. Therefore, the level of security here is always taken with respect to a fixed amount of resources. Once the first universal quantum computer is built, the RSA scheme will immediately become insecure. This is true for any encryption scheme for which there exists an efficient quantum algorithm that breaks it. Post-quantum algorithms, which are classical techniques that are believed to not be efficiently breakable by a quantum computer (see \cref{subsec:postquantum}), do not immediately become insecure once a quantum computer is built, but they still suffer from becoming gradually more and more insecure due to hardware and software developments. Consequently, which key length is considered secure in applications based on classical (including post-quantum) algorithms such as RSA has to be continuously updated. The security of quantum key distribution, on the other hand, does not change over time because it is information-theoretically secure and hence does not depend on how powerful state-of-the-art computers are. 

\section{Quantum cryptography within the larger cybersecurity ecosystem}
\label{sec:Quantumcrypto}

In contrast to classical cryptography, quantum cryptography can provide information-theoretic security that only relies on the laws of physics. In this section, we explain what this means, which role quantum cryptography can play within a larger cryptographic framework, and its limitations.

\subsection{Information-theoretic security}

What is required to guarantee a level of security that is independent of any computational assumptions, i.e., \emph{information-theoretic} or \emph{unconditional} security?  Before we attempt to answer this question, a note on this terminology is in order: Although the qualifier ``unconditional'' may suggest that no assumptions remain, certain assumptions about the underlying physical theory and the workings of the devices are still necessary (these will be discussed in \cref{subsec:assumptions}). In classical cryptography, these are usually implicit. Following this custom, we use the term as a synonym of ``information-theoretic security'' to mean that the computational assumptions are dropped completely.

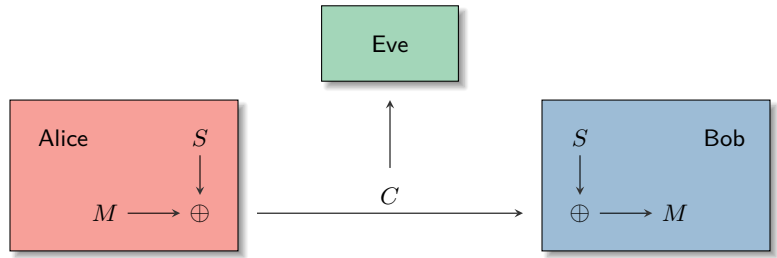
\begin{figure}[t]
	\centering
	\begin{tikzpicture}
		\draw[fill=mygreen!40,blur shadow={shadow blur steps=10}] (4.1,2.25) rectangle (5.9,3.25);
		\node at (5,2.75) {\small \textsf{Eve}};
		\draw[fill=myred!40,blur shadow={shadow blur steps=10}] (0,0) rectangle (3,2);
		\node at (0.7,1.5) {\small \textsf{Alice}};
		\node (oplusA) at (2.5,0.5) {$\oplus$};
		\node (SA) at (2.5,1.5) {\small $S$};
		\node (MA) at (1.25,0.5) {\small $M$};
		\draw[->,>=stealth] (MA) -- (oplusA);
		\draw[->,>=stealth] (SA) -- (oplusA);
		\draw[fill=myblue!40,blur shadow={shadow blur steps=10}] (7,0) rectangle (10,2);
		\node at (9.4,1.5) {\small \textsf{Bob}};
		\node (oplusB) at (7.5,0.5) {$\oplus$};
		\node (SB) at (7.5,1.5) {\small $S$};
		\node (MB) at (8.75,0.5) {\small $M$};
		\draw[->,>=stealth] (oplusB) -- (MB);
		\draw[->,>=stealth] (SB) -- (oplusB);
		\draw[->,>=stealth] (3.25,0.5) to node[above] {\small $C$} (6.75,0.5);
		\draw[->,>=stealth] (5,1.1) -- (5,2);
	\end{tikzpicture}
	\caption{\label{fig:onetimepad}\textbf{One-time pad encryption.} Alice and Bob share a cryptographic key $S$. Alice encrypts the message $M$ by performing bitwise addition modulo $2$ on the key and the message, $M\oplus S=C$, thus producing the ciphertext $C$. The ciphertext is then sent over a public channel to Bob. Bob uses the key to retrieve the message from the ciphertext by computing $C\oplus S=M$. An eavesdropper (Eve) only has access to the ciphertext $C$.}
\end{figure}

It is well understood how one can construct a cryptographic scheme for private communication that is secure independently of any hardness assumptions. The crucial ingredient here is a \emph{cryptographic key}, which is a sequence of uniformly random bits shared between two parties, whom we call Alice and Bob. To encrypt a message, Alice computes the bitwise addition modulo $2$ of the message and the key and sends the resulting ciphertext to Bob over a public channel. Bob can decrypt the ciphertext by adding the key to the ciphertext, thus retrieving the original message. Here, addition refers to bitwise addition modulo $2$, which is the same as subtracting the key from the ciphertext. This is known as the one-time pad (OTP) encryption method \cite{Vernam1926}, depicted in \cref{fig:onetimepad}. It is vital, though, that the key fulfills the following requirements:
\begin{enumerate}
	\item The key has to be truly random, which implies that the individual key bits cannot be correlated in any way.
	\item No part of the key can ever be reused. In particular, this implies that the key has to be at least as large as the entropy of the message that is supposed to be encrypted.
	\item It has to be securely delivered to Alice and Bob, who are possibly far apart, such that no one else has any information about the key.
\end{enumerate}
In this case, an eavesdropper who is keen to learn the message and has only access to the ciphertext and the general encryption method cannot learn anything about the encrypted message, which was shown by Claude Shannon in 1949 \cite{Shannon1949}. In addition to showing that the OTP protocol is information-theoretically secure, Shannon proved that any unbreakable encryption scheme must have the three characteristics listed above. 

Thus, the difficulty in constructing an unconditionally secure encryption system lies in generating a cryptographic key that fulfills the above conditions. This is where quantum mechanics comes into play. Quantum theory gives rise to several phenomena that do not exist in classical physics, which can be exploited in cryptography. They will be discussed in more detail in \cref{sec:Quantumtheory}. Because of these quantum mechanical properties, any attempt of an adversary to get information on the key leads to detectable changes in the system's state, which alert Alice and Bob of the adversary. The process of generating a cryptographic key using quantum theory is called \emph{quantum key distribution} (QKD).

Building on these properties, the security of quantum cryptography is based on the laws of physics rather than the conjectured hardness of certain problems. This makes it \emph{information-theoretically} secure, i.e., secure against an omnipotent adversary, which means that we not only consider them to have access to quantum computers but to any machine that might be developed in the future, regardless of how powerful it is (as long as it behaves according to the rules of quantum mechanics). This also implies that the security provided by QKD is \emph{future-proof} when used in combination with the one-time pad encryption, i.e., information encoded with it today will remain secure arbitrarily long, in contrast to an encoding using classical algorithms! In addition, the security of a quantum key distribution protocol can be quantified in the sense that we can give precise mathematical bounds on the probability that it can be cracked, which is explained in \cref{subsec:security}.

\subsection{Scope of applications for quantum cryptography}

With its guaranteed unconditional security, quantum key distribution (QKD) seems to be a promising candidate for replacing classical key generation schemes within larger cryptosystems. Nonetheless, this is only true to some extent. In general, we distinguish between two different kinds of encryption: In symmetric encryption the same key is used to encrypt and decrypt the information, whereas in asymmetric cryptography two keys are required, a public one and a private one. 

\subsubsection*{Symmetric encryption}

In principle, quantum key distribution is well suited to replace classical methods to generate symmetric keys. However, it is crucial to compose the quantum key distribution protocol with a suitable encryption method to preserve its security level. As explained above, combining a QKD protocol with one-time pad encryption provides information-theoretic security. This scheme, however, is somewhat unsuitable when it comes to practical terms because the key has to be at least as long as the message, and no part of it can be reused. Hence, if one wants to encrypt many messages, it is necessary to generate large amounts of secret key. Because the implementation of QKD protocols is, at the present time, considerably more expensive than classical key generation protocols, it is desirable to use the generated key as efficiently as possible.

Unfortunately, there is no real alternative to the one-time pad if one wants to maintain the unconditional security provided by the QKD protocol. One of the standard encryption protocols that are used nowadays is the Advanced Encryption Standard (AES) \cite{2001a}. The advantage of AES is that it requires a much shorter key to encrypt a given message compared to the one-time pad. The disadvantage, however, is that this encryption method only offers computational security, i.e., it is difficult to break in practice, but it does not provide the information-theoretic security that the one-time pad does. As a result, the composition of a QKD protocol and the AES encryption yields a scheme that is again just computationally secure. Still, it may be advantageous to provide fresh keys using QKD. For example, there might be a period of time where breaking AES is possible in principle, but it requires gathering a large amount of data to be successful. However, while this advantage can be justified heuristically, it cannot be proven or quantified.

\subsubsection*{Asymmetric encryption}

Asymmetric encryption schemes like RSA are widely used nowadays, for example, to secure communication between web browsers and to create digital signatures. As explained above, the RSA scheme relies on the conjecture that factoring large numbers is difficult on a classical computer. However, because Shor's algorithm provides a way to solve this problem on a quantum computer efficiently, RSA encryption will effectively be unsafe once quantum computers are built. Does quantum cryptography offer a possibility to replace classical asymmetric encryption methods with unconditionally secure quantum ones?

Unfortunately, there are no asymmetric quantum cryptographic protocols; the reason is that it is fundamentally impossible to achieve information-theoretic security of asymmetric encryption since one of the keys is always public. This prevents any such scheme from being unconditionally secure because the encryption of information with the public key can always, in principle, be reversed (it is necessary to not use irreversible encryption algorithms to ensure that the legitimate recipient can decrypt the message again). The security thus always relies on making the reversion process impractically hard, but if an adversary had unlimited computing power, they would be able to perform the reversion. Therefore, quantum key distribution is only suitable for the generation of symmetric keys, which is described in \cref{sec:QKDprotocols}.

\subsection{Post-quantum cryptography}
\label{subsec:postquantum}

With the looming advent of quantum computers and the accompanying threat of them being able to crack standard encryption methods, researchers have started to investigate classical techniques that are \emph{quantum safe}. This term refers to the fact that they cannot be broken efficiently by either a quantum or a conventional computer. Developing quantum-safe algorithms is all the more important because quantum cryptography cannot replace asymmetric classical encryption methods, as explained above. While certain asymmetric encryption schemes such as RSA immediately become insecure with the realization of a universal quantum computer, symmetric encryption schemes such as AES are not as much at risk. They are thought to be most vulnerable to quantum brute force attacks enabled by Grover's algorithm \cite{Grover1996}, which can provide a quadratic speedup over the best known classical algorithms. An efficient countermeasure against these attacks is thus increasing the length of the keys used to ensure security. For example, the AES scheme with $256$ bit keys is currently considered quantum-safe because, considering brute force attacks, it is as difficult to break for a quantum computer as the $128$ bit key is for a classical computer. 

The goal of post-quantum cryptography \cite{Bernstein2009,Bernstein2017} is thus to find algorithms that are hard for quantum computers (which, of course, includes classical computers as a special case). In some aspects, it faces similar problems as ``traditional'' classical cryptography: The security of post-quantum cryptographic schemes relies on the assumption that the underlying problem is hard to crack for quantum computers. Consequently, the question of which mathematical problems are considered quantum-safe needs to be updated regularly as quantum computers become more powerful and new quantum algorithms are developed. However, when it comes to the question of which algorithms are difficult for a quantum computer, researchers have the disadvantage that they cannot rely on decade-long trials, as is the case with many problems considered difficult to solve on classical computers. In other words, we do not really know yet which problems are hard for quantum computers, for the search for quantum algorithms to solve them only started quite recently.

{\renewcommand{\arraystretch}{1.4}\setlength{\tabcolsep}{5pt} 
	\begin{table}[t]
		\centering
		\begin{tabular}{r"C{2.7cm}|C{2.7cm}|C{2.7cm}}
			& Hardware & Software & Developments in \\[-0.5em]
			& developments & developments & physical theories \\\thickhline
			Classical cryptography & \cellcolor{myred!30} not safe & \cellcolor{myred!30} not safe & \cellcolor{myred!30} not safe\\\hline
			Post-quantum cryptography & \cellcolor{mygreen!30} safe & \cellcolor{myred!30} not safe & \cellcolor{myred!30} not safe \\\hline
			Quantum cryptography & \cellcolor{mygreen!30} safe & \cellcolor{mygreen!30} safe & \cellcolor{myred!30} not safe\\\hline
			Non-signaling cryptography & \cellcolor{mygreen!30} safe & \cellcolor{mygreen!30} safe & \cellcolor{mygreen!30} safe
		\end{tabular}
		\caption{\label{tab:postcrypto}\textbf{Possible threats for cryptography.} Classical, post-quantum, quantum, and non-signaling cryptography exhibit different levels of security with regard to threats concerning hardware and software developments and new insights into the underlying physical theories. In the table, ``not safe'' means that there is no guarantee that the respective cryptographic scheme will remain secure under the listed developments.}
	\end{table}
}
%

Despite these difficulties, post-quantum cryptography has a crucial advantage compared to classical cryptography: It does not have to fear the development of a more powerful machine because a universal quantum computer is already the worst-case scenario, at least within the theory of quantum mechanics. The gradual improvement of hardware might make it necessary to adjust the length of the keys used in post-quantum algorithms (as explained in \cref{subsec:quantitative}) from time to time, but the development of a universal quantum computer will not make them immediately insecure.

\subsection{Cryptography beyond quantum theory}

It is of course possible that, in addition to the development of hardware and software, new findings regarding the physical theories that describe our universe will also affect cryptography. For instance, quantum theory will likely have to be modified to allow for a theory that unifies quantum mechanics and gravity. This development would then not only have an impact on classical and post-quantum cryptography, but also quantum cryptography, as it relies on quantum theory. The changes that have to be made to quantum theory, however, will not affect each part of the theory equally. As long as the quantum effects responsible for the security of quantum cryptography (\cref{sec:Quantumtheory}) remain roughly the same, the security of quantum cryptography is not at risk. In a theory of quantum gravity, these effects hence have to be revisited again. We can even go one step further and formulate cryptographic protocols that do not depend on the correctness of quantum theory, but only rely on the non-signaling condition, which means it is only required that faster-than-light communication is impossible \cite{Barrett2005}. This kind of protocol is even secure against developments in physical theories, as we expect any theory that describes our universe to fulfill the non-signaling condition. The vulnerabilities of the different approaches to cryptography to potential threats are summarized in \cref{tab:postcrypto}.

\section{Concepts of quantum theory for cryptography}
\label{sec:Quantumtheory}

In the last section, we have seen that quantum mechanics can guarantee unconditional security due to the fact that it relies on the laws of physics rather than the presumed computational hardness of certain problems. But what is it that makes cryptography based on quantum mechanics so much more secure than its classical counterpart? In this section, we explain the underlying concepts of quantum physics and show how they play together to guarantee unconditional security.

\subsection{The state space of quantum mechanics}
\label{subsec:statespace}

When thinking of quantum mechanics, the first thing that comes to mind is usually the Schrödinger equation, which describes the time evolution of a quantum system in terms of its wave function. Although this is undoubtedly one of the most famous elements of quantum theory, it is not part of the fundamental concepts that we use in quantum cryptography. Instead, we exploit the specific structure and properties of the state space of quantum mechanics, which gives rise to unique quantum phenomena such as superposition, entanglement, and uncertainty. 

\subsubsection*{Superposition}

To illustrate these quantum phenomena, let us consider an example. An electron is a subatomic particle that can be in a state where its position is not uniquely determined. If it is part of a hydrogen atom, it orbits a positively charged nucleus at some distance, bound by the Coulomb force. According to quantum mechanics, the exact position of the electron cannot be predicted because this property of the electron does not have a predetermined value. If we nevertheless carry out a position measurement on the electron, we will get a specific position as a result. This result, however, has been produced randomly according to some probability distribution; it didn't exist prior to the measurement. The electron example shows one of the fundamental properties of quantum theory, called indeterminism. It describes the fact that properties of quantum objects such as position, momentum, and energy do not necessarily have a definite value. We can only assign probabilities to the different values of these properties.

\begin{figure}[t]
	\centering
	\begin{tikzpicture}[scale=1.1,domain = 0:3.5, samples = 50]
		\draw[fill=LightGray,draw=black] (-4,-0.35) rectangle (-2.5,0.35);
		\node at (-3.25,0.15) {\footnotesize electron};
		\node at (-3.25,-0.15) {\footnotesize gun};
		\foreach \a in {0,20,40,-20,-40}: 
		\draw[->,>=stealth] (-2.5,0) -- ($(-2.5,0)+(\a:0.75)$);
		\draw[dashed] (-1.75,0) -- (2.35,0);
		\draw[fill=gray,draw=black] (-0.15,0.75) rectangle (0.15,1.75);
		\draw[fill=gray,draw=black] (-0.15,-0.5) rectangle (0.15,0.5);
		\draw[fill=gray,draw=black] (-0.15,-0.75) rectangle (0.15,-1.75);
		\node[color=myred] at (-0.4,0.75) {\small $S_1$};
		\node[color=myblue] at (-0.4,-0.5) {\small $S_2$};
		\draw[fill=gray,draw=black] (2.35,-1.75) rectangle (2.65,1.75);
		\node at (2.1,1.7) {\small $D$};
		\draw[->,>=stealth] (3.5,-1.75) -- (3.5,1.75);
		\draw[->,>=stealth] (3.5,0) -- (5,0);
		\node at (3.5,1.9) {\small $x$};
		\node at (5.15,0) {\small $P$};
		\node[color=myred] at (4.25,1.15) {\small $P_1$};
		\node[color=myblue] at (4.25,-1.15) {\small $P_2$};
		\begin{scope}[xshift=3.5cm,yshift=1.75cm]
			\draw[rotate around={-90:(0,0)},draw=myred, fill=myred, fill opacity=0.1, thick, smooth] plot (\x,{gauss(1.35,0.4)});
			\draw[rotate around={-90:(0,0)},draw=myblue, fill=myblue, fill opacity=0.1, thick, smooth] plot (\x,{gauss(2.15,0.4)});
		\end{scope}
		\draw[->,>=stealth] (6,-1.75) -- (6,1.75);
		\draw[->,>=stealth] (6,0) -- (7.5,0);
		\node at (6,1.9) {\small $x$};
		\node at (7.65,0) {\small $P$};	
		\node[color=mygreen] at (6.75,1.15) {\small $P_{12}$};
		\begin{scope}[xshift=6.01cm,yshift=1.75cm]
			\draw[rotate around={-90:(0,0)},draw=mygreen, fill=mygreen, fill opacity=0.1, thick, smooth] plot (\x,{gauss(1.75,0.5)*newcos});
		\end{scope}
	\end{tikzpicture}
	\caption{\label{fig:doubleslit}\textbf{Double-slit experiment with electrons.} The electron gun sends electrons towards a wall with two slits in it, $S_1$ and $S_2$. Some electrons travel through the slits towards a detector $D$, for example, an electron multiplier that makes a sound whenever it detects an electron. We can use this setup to collect statistics about where the electrons are detected and plot the corresponding probability distributions. Whenever only one of the two slits is open, we see either the probability distribution $P_1$ or $P_2$ (depending on which slit is open). When both slits are open we observe the probability distribution $P_{12}$. From the diagram it becomes clear that $P_{12}\neq P_1+P_2$, showing that the electron does not simply pass through one of the possible slits, but interferes with itself (which corresponds to being in a superposition of the two possibilities).}
\end{figure}
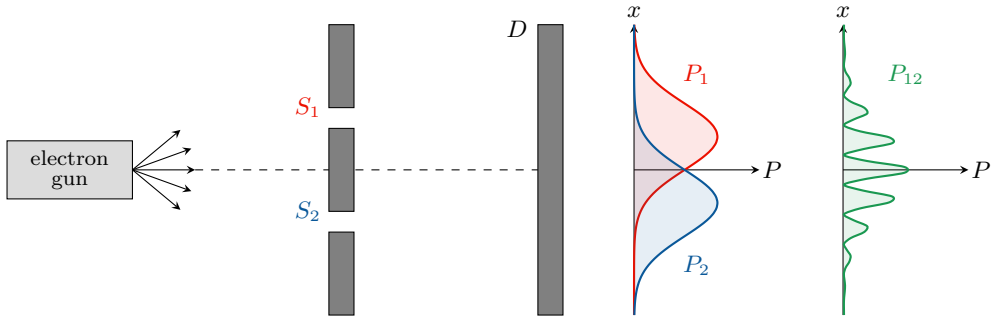

Formally, the phenomenon that the electron does not have a definite position is described by the concept of superposition. Consider the situation in a double-slit experiment as depicted in \cref{fig:doubleslit}. The electron can either pass through the first slit or the second. Moreover, it can also be in a superposition of passing through the first and the second slit. This is sometimes described as the electron being in both positions at once, i.e., it takes both possible paths. However, this does not capture the situation correctly: We can only say that the probability of finding the electron in either path upon measuring it is nonzero, but a measurement would always either output the result ``electron in path $1$'' or ``electron in path $2$'', but never the result ``electron in path $1$ \emph{and} $2$''. This behavior is verified in the experiment by the interference pattern detected behind the double-slit that occurs if we only measure where the electron impinges on the screen, but not which path it has taken. If, on the other hand, we measure the position of the electron when passing the double-slit, we will randomly get one of the two positions, slit one or slit two. The measurement forces the electron to decide which path it takes. The concept of quantum superposition has no analog in classical physics; it is essentially different from anything we can observe in any classical theory.

The electron in the double-slit experiment is an example of a quantum system that is in a superposition of two states, namely the state ``passes through slit $S_1$'' and the state ``passes through slit $S_2$''. As such, it is a realization of a \emph{quantum bit} or \emph{qubit}, for short. This is a generalization of a classical bit and commonly used in quantum cryptography. The two values, corresponding to two perfectly distinguishable quantum states of the system, are typically denoted $\ket{0}$ and $\ket{1}$. As explained above, a quantum system is not necessarily in one of these possible states but can be in a superposition of them. This means that the general state of a qubit can be written as a linear combination of the two possible classical states:
\begin{equation}
	\label{eq:qubitcomp}
	\ket{\psi}=\alpha\ket{0}+\beta\ket{1}.
\end{equation}
Here, we see a first difference between bits and qubits: While classical bits take either the value $0$ or $1$, quantum bits can be in a superposition of the two states $\ket{0}$ and $\ket{1}$, and only upon measuring them they have to choose one. This is analogous to the electron in the double-slit experiment, whose path is in a superposition of both slits until we measure which path it takes.
From the coefficients $\alpha$ and $\beta$, we can calculate the probability of getting the result $0$ or $1$, respectively, when measuring the qubit:
\begin{equation}
	\label{eq:probrule}
	\mathrm{Pr}\big[0\big]=|\alpha|^2,\hspace{10pt} \mathrm{Pr}\big[1\big]=|\beta|^2,
\end{equation}
where $\alpha$ and $\beta$ are complex numbers. Because the probability of finding the object in any one of the possible states has to be $1$, it follows that $|\alpha|^2+|\beta|^2=1$. If both results, $0$ and $1$, occur with equal probability upon measuring the qubit, $\alpha$ and $\beta$ must have the same value, hence the state could be
\begin{equation}
	\ket{\psi}=\frac{1}{\sqrt{2}}\ket{0}+\frac{1}{\sqrt{2}}\ket{1}.
\end{equation}

A priori, a qubit is an abstract notion that does not specify the physical system we are working with. A possible realization  of a qubit (apart from the electron in a double-slit experiment) is provided by polarized photons. For example, we can identify the state $\ket{0}$ with the horizontal polarization and the state $\ket{1}$ with the vertical polarization. However, these are not the only choices for realizing the two states via directions of polarization. Alternatively, we could choose to identify the state $\ket{0}$ with a diagonal polarization at angle $+45^\circ$ and the state $\ket{1}$ with one at angle $-45^\circ$. To distinguish between these two choices, we can use a different notation for the diagonally polarized states: We denote polarization at angle $+45^\circ$ with the state $\ket{+}$, and polarization at angle $-45^\circ$ with $\ket{-}$. Analogous to \cref{eq:qubitcomp}, we can hence realize a qubit with the two polarization states $\ket{+}$ and $\ket{-}$, whose general state is
\begin{equation}
	\ket{\psi}=\alpha'\ket{+}+\beta'\ket{-}.
\end{equation}

The two pairs of directions of polarization we have introduced above, $\{\ket{0},\ket{1}\}$ and $\{\ket{+},\ket{-}\}$, are not independent of each other. They are connected via the relations
\begin{equation}
	\label{eq:qubitpm}
	\ket{+}=\frac{1}{\sqrt{2}}\ket{0}+\frac{1}{\sqrt{2}}\ket{1},\hspace{10pt} \ket{-}=\frac{1}{\sqrt{2}}\ket{0}-\frac{1}{\sqrt{2}}\ket{1}.
\end{equation}

\tikzset{color1/.style = {color=myred}}
\tikzset{color2/.style = {color=myblue}}

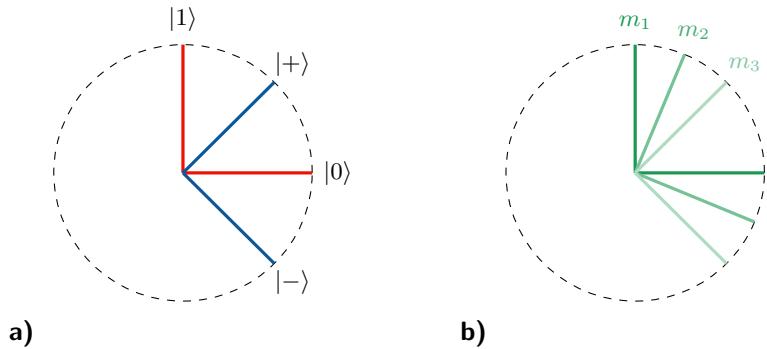
\begin{figure}
	\centering
	\begin{tikzpicture}[scale=0.85]
		\node at (-2.5,-2.5) {\textbf{\textsf{a)}}};
		\draw[dashed] (0,0) circle (2cm);
		\draw[color1, very thick] (0,0) -- (0,2);
		\draw[color1, very thick] (0,0) -- (2,0);
		\draw[color2, very thick] (0,0) -- (45:2);
		\draw[color2, very thick] (0,0) -- (315:2);
		\node at (0:2.4cm) {\small $\ket{0}$};
		\node at (90:2.4cm) {\small $\ket{1}$};
		\node at (45:2.4cm) {\small $\ket{+}$};
		\node at (315:2.4cm) {\small $\ket{-}$};
		\begin{scope}[xshift=7cm]
			\node at (-2.5,-2.5) {\textbf{\textsf{b)}}};
			\draw[dashed] (0,0) circle (2cm);
			\foreach \a in {0,22.5,45,67.5,90}{
				
			}
			\draw[gray, very thick, color=mygreen]  (0,0) -- (0:2);
			\draw[gray, very thick, color=mygreen]  (0,0) -- (90:2);
			\draw[gray, very thick, color=mygreen!60]  (0,0) -- (67.5:2);
			\draw[gray, very thick, color=mygreen!60]  (0,0) -- (-22.5:2);
			\draw[gray, very thick, color=mygreen!35]  (0,0) -- (45:2);
			\draw[gray, very thick, color=mygreen!35]  (0,0) -- (-45:2);
			\node[color=mygreen] at (90:2.35cm) {\small $m_1$};
			\node[color=mygreen!80] at (67.5:2.375cm) {\small $m_2$};
			\node[color=mygreen!55] at (45:2.4cm) {\small $m_3$};
		\end{scope}
	\end{tikzpicture}
	\caption{\label{fig:qubitpol}a)~Possible realizations of a qubit via photon polarization. The red lines visualize the states $\ket{0}$ and $\ket{1}$, and the blue lines visualize the states $\ket{+}$ and $\ket{-}$. b)~Some possible measurements $m_i$ of photon polarization. The more parallel two sets of lines are, the more similar are the corresponding probability distributions of their outcomes. Note that $m_1$ corresponds to a measurement in the horizontal/vertical polarization basis, and $m_3$ corresponds to measurement in the diagonal polarization basis.}
\end{figure}

The two possible ways of using polarization of photons are illustrated in \cref{fig:qubitpol}a. Horizontal polarization, denoted $\ket{0}$, is depicted as a horizontal red line. Vertical polarization $\ket{1}$ accordingly corresponds to a horizontal red line. The states $\ket{+}$ and $\ket{-}$ correspond to blue lines at angles $+45^\circ$ (with respect to the vertical red line) and $-45^\circ$, respectively.

The measurement of the polarization of a photon can be depicted in a similar way, see \cref{fig:qubitpol}b. For example, the measurement $m_1$, which is illustrated by vertical and a horizontal line (similar to the states $\ket{0}$ and $\ket{1}$ in \cref{fig:qubitpol}a), corresponds to a measurement that can distinguish horizontal and vertical polarization. We say that it is a measurement in the $\{\ket{0},\ket{1}\}$-basis. If the photon is horizontally polarized (which corresponds to the state $\ket{0}$), such a measurement will always yield the result $0$. If, instead, the photon is diagonally polarized at angle $+45^\circ$ (i.e., in the state $\ket{+}$), we will get the results $0$ and $1$ each with a probability of $50\%$ because of the relation \cref{eq:qubitpm} and the rule for calculating probabilities given in \cref{eq:probrule}. If we interpret the two possible measurement outcomes as the two values a classical bit can take, the above observations show how measuring quantum states in a superposition allows us to generate truly random bits.


We can make a similar observation when instead of changing the state, we change the measurement. What happens when we measure the state $\ket{0}$ with the measurement direction $m_3$, i.e., the one that is parallel to the states $\ket{+}$ and $\ket{-}$? In this case, we will also get the two possible results with equal probability. If we instead choose the measurement $m_2$, whose lines are \emph{almost} parallel to the red lines, to measure the state $\ket{0}$, we will get the result $0$ in a majority of the cases, i.e., we get results which are \emph{almost} the same as the ones for $m_1$.

From these observations, we can derive some general rules: If one of the lines that represent the measurement is parallel to the line that represent the state, the measurement outcome will always be the same (as the measurement can perfectly distinguish between the two states). Furthermore, the more parallel the lines of two different measurements arrows are, the more similar are the respective probability distributions of the results upon measuring the same state.

\subsubsection*{Entanglement}

A single quantum particle already shows remarkable features that cannot be observed in classical systems. However, quantum particles do not exist isolated. Several quantum particles can be composed to form a bigger quantum system. These kind of composed quantum systems exhibit an additional feature that is unique to quantum theory, namely \emph{entanglement}.

Let us first briefly go back to the electron example from above. We have seen that an electron does not have to be in a definite position but can be in a superposition of different positions. We can now form a composite system of two electrons. Similar to single-particle systems, a composite system of two electrons can also be in a superposition. However (and here the full extent of the strangeness of quantum mechanics reveals itself!), it is possible that even though neither electron has a definite position, the distance between them is well-defined. In this case, we say that the two electrons are entangled.

To explain the use of entanglement, we now focus on qubits: Suppose two parties, Alice and Bob, share an entangled state of the form
\begin{equation}
	\label{eq:entangledstate}
	\ket{\Phi}_{AB}=\frac{\ket{0}_A\ket{0}_B+\ket{1}_A\ket{1}_B}{\sqrt{2}},
\end{equation}
where the qubit with index $A$ belongs to Alice and the one with index $B$ belongs to Bob. This state is called a \emph{Bell state}. In this case, Alice and Bob do not know whether the individual qubits are in state $\ket{0}$ or $\ket{1}$, but they do know that both qubits are in the same state (because no terms of the form $\ket{0}_A\ket{1}_B$ or $\ket{1}_A\ket{0}_B$ appear). The same is true if we consider the diagonal polarization states $\ket{+}$ and $\ket{-}$, i.e., using \cref{eq:qubitpm} one can show that the Bell state can be expressed equivalently as 
\begin{equation}
	\ket{\Phi}_{AB}=\frac{\ket{+}_A\ket{+}_B+\ket{-}_A\ket{-}_B}{\sqrt{2}}.
\end{equation}
That means that whenever Alice and Bob measure their individual qubits in the same basis, they will obtain the same results. As such, shared entanglement can be used as a resource in quantum cryptography, where Alice and Bob use the outcomes of their measurements to produce a shared secret key. Suppose Alice and Bob share a number of entangled states $\ket{\Phi}_{AB}$ and measure their respective parts in order to generate a key. As long as their respective measurement bases are always the same, they will obtain perfectly correlated outcomes. If they choose measurement bases that are conjugate to each other, for example Alice measures in the $\{\ket{0},\ket{1}\}$-basis and Bob in the $\{\ket{+},\ket{-}\}$-basis, their outcomes are uncorrelated. Thus, after the measurement step it is necessary that Alice and Bob exchange the information in which basis they have measured each state and discard the uncorrelated outcomes. Note that even though Alice and Bob get correlated results, these results are still perfectly random, and in particular independent of the choice of measurement basis. Hence, publicly revealing the chosen bases does not corrupt the secrecy of the generated bits.

\subsection{Information gain vs.\ state disturbance}

The structure of quantum states described above has a remarkable consequence for the ability to gain information about a system. Consider the scenario described above, where Alice and Bob share a number of entangled states and measure them in either the $\{\ket{0},\ket{1}\}$- or the $\{\ket{+},\ket{-}\}$-basis to generate a shared secret key. Let us now add a third party to the setting, an eavesdropper, called Eve, who wants to gain as much information as possible without being detected. If the eavesdropper is detected, Alice and Bob will simply abort the protocol and the eavesdropper does not gain any usable information. In quantum cryptographic settings, the entangled states are typically produced either outside Alice and Bob's labs or in one of the parties' lab and sent to one or both of them via insecure quantum channels. One possible way of gaining information for Eve is to intercept the quantum states and measure them before they reach Alice's and Bob's respective laboratories.

For instance, suppose that Alice produces the entangled states in her lab and sends one half of each state to Bob. Eve can intercept the part that is sent to Bob, measure it, and resend the state to Bob. However, there are two problems here that limit the amount of information Eve can obtain: First, because she does not know in which basis Alice will measure her qubit, she has to guess the measurement basis. If she chooses a different one than Alice, the results are uncorrelated and Eve does not gain any information. Second, upon measuring the qubit that is supposed to be sent to Bob, Eve inevitably disturbs its quantum state if she wants to gain information. This stems from the fact that every informative measurement disturbs the state of the system \cite{Bennett1992a,Busch2009}. In fact, there is a trade-off between the information Eve can learn and the extent to which the state is altered: the more information she gains by measuring, the more she changes the state, which increases the chance that her attack will be detected.

Is it possible for Eve to build a measurement apparatus that measures both quantities at once, polarization in the $\{\ket{0},\ket{1}\}$-basis and in the $\{\ket{+},\ket{-}\}$-basis? This would solve the problem that the outcomes are uncorrelated if Eve chooses a different basis than Alice because she does not have to make a choice anymore. However, in quantum mechanics such apparatuses exist only for a limited set of \emph{jointly measurable} or \emph{compatible} variables, which is a consequence of \emph{Heisenberg's uncertainty principle} \cite{Heisenberg1927}. The two polarization bases do not fall into this set. When generating a key, Alice and Bob will always choose variables that are not jointly measurable, hence it is impossible for Eve to measure both quantities simultaneously. 
In classical physics, it is generally not a problem to measure two properties of an object simultaneously. This is an important difference between classical and quantum physics.

\subsection{Quantum cloning is impossible}
\label{subsec:nocloning}

To gain information about a state in a quantum key generation procedure Eve cannot measure all properties of the quantum state simultaneously, as discussed above. Is it possible that Eve simply copies the state that is sent to Bob? She could then forward the original, undisturbed, state to Bob and perform measurements on her copy of the state. She could even produce several copies of the state in order to perform different measurements.

\begin{figure}[t]
	\centering
	\begin{tikzpicture}
		\begin{scope}[xscale=-1,xshift=-2.5cm]
			\draw[fill=myblue!40] (2.3,0.8) to[bend left] (2.3,1.1) -- (2.7,1.3) to[bend left] (2.7,0.6) -- cycle;
		\end{scope}
		\draw[fill=myblue!30,draw=white] (0,0) rectangle (2,1.5);
		\draw[fill=myblue!50,draw=white] (0,1.5) -- (0.5,2) -- (2.5,2) -- (2,1.5);
		\draw[fill=myblue!50,draw=white] (2,1.5) -- (2.5,2) -- (2.5,0.5) -- (2,0);
		\draw (0,1.5) -- (0,0) -- (2,0) -- (2,1.5) -- (0,1.5) -- (0.5,2) -- (2.5,2) -- (2,1.5) -- (2,0) -- (2.5,0.5) -- (2.5,2);
		\draw[fill=myblue!40] (2.3,0.8) to[bend left] (2.3,1.1) -- (2.7,1.3) -- (2.7,0.6) -- cycle;
		\draw[fill=myblue!60] (2.7,1.3) to[bend left] (2.7,0.6) to[bend left] (2.7,1.3);
		\node at (1,0.95) {\small \texttt{Quantum}};
		\node at (1,0.55) {\small \texttt{Cloner}};
		\node (onepsi) at (-1.5,1) {$|\psi\rangle$};
		\node (twopsi) at (4.25,1) {$|\psi\rangle|\psi\rangle$};
		\draw[->,>=stealth,thick] (onepsi) -- (-0.5,1);
		\draw[->,>=stealth,thick] (3,1) -- (twopsi);
	\end{tikzpicture}
	\caption{\label{fig:Nocloning}\textbf{A universal cloning machine.} This machine is able to copy an arbitrary, unknown quantum state $|\psi\rangle$ such that we get two perfect copies of the state. Its existence is forbidden by the laws of quantum mechanics.}
\end{figure}
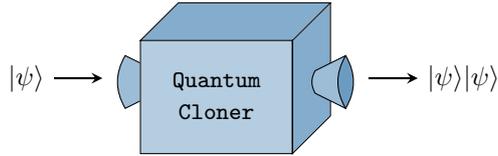

Fortunately, the structure of quantum mechanics implies that \emph{perfectly cloning an arbitrary unknown quantum state is impossible} \cite{Wootters1982,Dieks1982}. Why can we not make a copy of a quantum state? Let us see what happens if we assume that it is possible to build a machine that can perfectly clone arbitrary quantum states, as illustrated in \cref{fig:Nocloning}. In the following, we will show via an argument by contradiction that having access to such a machine would allow faster-than-light communication, which would violate special relativity. Ergo, we can conclude that it must be impossible to clone arbitrary quantum states. 

The argumentation is as follows: Consider a scenario where Alice and Bob share a Bell state $\ket{\Phi}_{AB}$, which is an entangled two-qubit state that can be written either in the $\{\ket{0},\ket{1}\}$- or the $\{\ket{+},\ket{-}\}$-basis:
\begin{equation}
	\label{eq:Bell}
	\ket{\Phi}_{AB}=\frac{\ket{0}_A\ket{0}_B+\ket{1}_A\ket{1}_B}{\sqrt{2}}=\frac{\ket{+}_A\ket{+}_B+\ket{-}_A\ket{-}_B}{\sqrt{2}},
\end{equation}
where the first qubit belongs to Alice and the second one to Bob. If Alice measures her part of the state in the $\{\ket{0},\ket{1}\}$-basis and gets either $0$ or $1$ as the result, Bob's qubit will be in the corresponding state afterwards. If he then performs a measurement in the $\{\ket{0},\ket{1}\}$-basis on his qubit he will get the same result as Alice with certainty. Similar for the $\{\ket{+},\ket{-}\}$-basis: Whether Alice gets the result $+$ or $-$, Bob's qubit will be in the respective state afterwards. However, Bob is unable to learn which basis Alice has measured in just by measuring his qubit. A single measurement outcome does not reveal anything about the probability distribution of the outcomes.

Suppose now that Bob has access to the quantum cloning machine depicted in \cref{fig:Nocloning} and he uses it to produce $1,000$ perfect copies of his state. After Alice has measured her part of the state Bob measures half of them in the $\{\ket{0},\ket{1}\}$-basis and the other half in the $\{\ket{+},\ket{-}\}$-basis. Say, Alice has measured her qubit in the $\{\ket{0},\ket{1}\}$-basis and has gotten $0$ as a result. The states that Bob measures in the $\{\ket{0},\ket{1}\}$-basis will then produce the output $0$ with certainty, while the states that are measures in the $\{\ket{+},\ket{-}\}$-basis will yield random outputs. As a result, Bob will observe (roughly) the following probability distribution:
\begin{table}[H]
	\centering
	\begin{tabular}{rl}
		$\{\ket{0},\ket{1}\}$-basis:& $\mathrm{Pr}\big[0\big]=100\%$ \\[2pt]
		& $\mathrm{Pr}\big[1\big]=0\%$ \\[5pt]
		$\{\ket{+},\ket{-}\}$-basis:& $\mathrm{Pr}\big[+\big]=50\%$ \\[2pt]
		& $\mathrm{Pr}\big[-\big]=50\%$
	\end{tabular}
\end{table}

From this probability distribution Bob can deduce that Alice has measured in the $\{\ket{0},\ket{1}\}$-basis, which corresponds to the superluminal transmission of one bit (given that they are far apart, but because this argumentation holds for any distance between Alice and Bob we can make this distance sufficiently large). However, because we have assumed that faster-than-light communication is impossible, it follows that perfect cloning of arbitrary quantum states must be impossible.

The impossibility of quantum cloning directly implies that one cannot get information about a quantum state without disturbing it (which was explained in the previous section). If all properties of the quantum state could be measured without changing the state, we had perfect knowledge about it and hence could generate another, identical, quantum state while the first one still exists.

\subsection{Monogamy of entanglement}
\label{subsec:monogamy}

So far, we have seen that an eavesdropper can neither gain information by measuring all properties of a quantum system simultaneously nor can she make copies of the states that are sent. Maybe it is possible for Eve instead to entangle her quantum system with Alice's quantum state such that her measurement results are perfectly correlated with Alice's measurement outcomes? In other words, is it possible that Alice's quantum system is perfectly entangled both with Bob's quantum state and with Eve's? In this way, Eve would get the maximum amount of information without being detected because she never has to perform any measurement on Alice's and Bob's respective quantum states.

Fortunately (for the sake of quantum cryptography), this strategy is prohibited by what is called the \emph{monogamy of entanglement} \cite{Coffman2000}. This term describes the phenomenon that when two quantum systems are perfectly entangled (such as Alice's and Bob's quantum systems are if they are in the state \cref{eq:Bell}), neither of them can be entangled to a third system. This follows directly from the fact that quantum cloning is impossible, which was shown in the preceding section. Or, put differently: If ``polygamy'' of entanglement were possible, cloning arbitrary unknown quantum states would be possible. 

\begin{figure}[t]
	\centering
	\begin{tikzpicture}
		\begin{scope}[yshift=0.975cm]
			\node at (-0.625,3) {\color{myred} \textsf{Alice}};
			\draw[fill=myred!40,blur shadow={shadow blur steps=10}] (-1.1,0) rectangle (1.1,2.75);
			\node (BellA) at (-0.25,2.25) {$\bullet$};
			\node at (0.25,3.5) {\small $\ket{\psi}$};
			\draw[fill=LightGray] (-0.5,1.1) rectangle (0.5,1.65);
			\node at (0,1.375) {\small $\mathcal{B}$};
			\draw[->,>=stealth] (-0.25,2.1) -- (-0.25,1.75);
			\draw[->,>=stealth] (0.25,3.25) -- (0.25,1.75);
			\draw[->,>=stealth] (0,1) -- (0,0.65);
			\node at (0,0.4) {\small $x,y$};
		\end{scope}
		\node at (5.65,3) {\color{myblue} \textsf{Bob}};
		\draw[fill=myblue!40,blur shadow={shadow blur steps=10}] (3.9,0) rectangle (6.1,2.75);
		\node (BellB) at (5,2.25) {$\bullet$};
		\draw[->,>=stealth] (5,2.1) -- (5,1.75);
		\draw[fill=LightGray] (4.5,1.1) rectangle (5.5,1.65);
		\node at (5,1.375) {\small $\mathcal{M}$};
		\draw[->,>=stealth] (5,1) -- (5,-0.5);
		\node at (5,-0.75) {\small $\ket{\psi}$};
		\node at (-5.5,3) {\color{mygreen} \textsf{Charlie}};
		\draw[fill=mygreen!40,blur shadow={shadow blur steps=10}] (-6.1,0) rectangle (-3.9,2.75);
		\node (BellC) at (-5,2.25) {$\bullet$};
		\draw[->,>=stealth] (-5,2.1) -- (-5,1.75);
		\draw[fill=LightGray] (-4.5,1.1) rectangle (-5.5,1.65);
		\node at (-5,1.375) {\small $\mathcal{M}$};
		\draw[->,>=stealth] (-5,1) -- (-5,-0.5);
		\node at (-5,-0.75) {\small $\ket{\psi}$};
		\draw[->,>=stealth] (0.5,1.375) to node[above] {\footnotesize \text{classical channel}} (4.35,1.375);
		\draw[->,>=stealth] (-0.5,1.375) to node[above] {\footnotesize \text{classical channel}} (-4.35,1.375);
		\draw[color=gray!70!black,snake=snake] (BellA) -- (BellB);
		\node[color=gray!70!black] at (2.5,3) {\rotatebox{-12}{\footnotesize entangled state}};
		\draw[color=gray!70!black,snake=snake] (BellA) -- (BellC);
		\node[color=gray!70!black] at (-2.5,3.05) {\rotatebox{12}{\footnotesize entangled state}};
	\end{tikzpicture}
	\caption{\label{fig:Qteleport}\textbf{Quantum cloning via teleportation.} We can show the monogamy of entanglement via an argument by contradiction: Suppose that Alice possesses a qubit state that is perfectly entangled with both Bob's and Charlie's respective quantum states. She performs a Bell measurement $\mathcal{B}$ on this state and the state $\ket{\psi}$ she wants to clone. This measurement outputs two classical bits $x$ and $y$, which Alice sends to Bob and Charlie over classical channels. The values of the two bits provide the information which transformation $\mathcal{M}$ Bob and Charlie have to perform on their respective states in order to transform those to the state $\ket{\psi}$. This scheme effectively copies the qubit state $\ket{\psi}$ without requiring any knowledge of the state itself. Because we know that cloning is impossible, this argument shows that polygamy of entanglement is impossible.}
\end{figure}
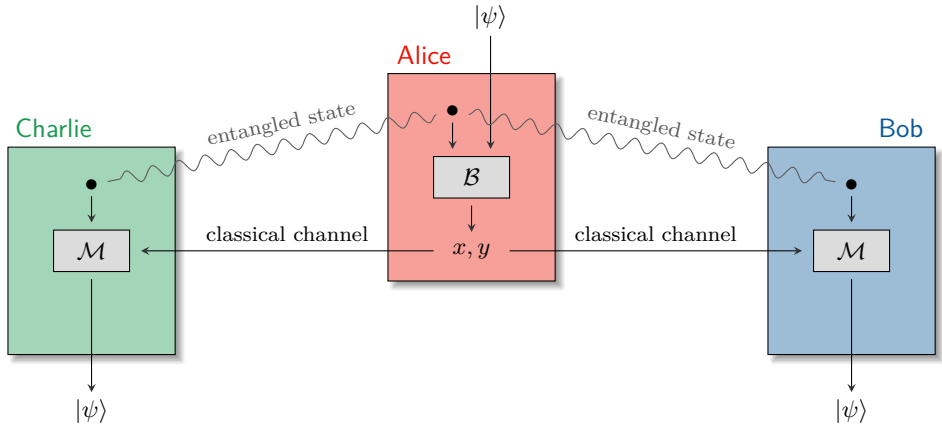

This can be illustrated using a protocol called \emph{quantum teleportation} \cite{Bennett1993}, whose goal is to send a quantum state from one party to another (spatially distant) party without using a quantum channel. However, the two parties are allowed to exchange classical information. It works as follows: Alice and Bob share a perfectly entangled state (for example, the Bell state in \cref{eq:Bell}). To teleport an arbitrary qubit state $\ket{\psi}$ to Bob, Alice measures her part of the entangled state together with the state $\ket{\psi}$ in a Bell measurement. This measurement effectively outputs two classical bits $x$ and $y$. Alice then sends these bits to Bob via a classical channel. The bits provide the information which transformation Bob has to perform on his part of the entangled state to transform it into the state $\ket{\psi}$. Note that this does not violate the fact that quantum states cannot be cloned: The state $\ket{\psi}$ is destroyed at Alice's lab by the Bell measurement before it is recreated in Bob's lab. This, together with the fact that the information never has to transit between Alice and Bob, is responsible for the name ``teleportation''. Additionally, quantum teleportation does not violate the no-signaling principle: Before Bob applies the transformation his state does not contain any information, and the information which transformation to apply cannot be transmitted faster than light.

Bob only knows which transformation he has to perform after receiving the two classical bits Alice sends to him, which cannot be transmitted faster than light.

Suppose it were possible for Alice to be entangled with more than one party. In that case, she could execute the same protocol simultaneously with Bob and a third party, called Charlie (see \cref{fig:Qteleport}). First, she entangles her quantum state with both Bob's and Charlie's respective quantum states. She then continues with the protocol steps as described above but sends the results $x,y$ of the Bell measurement to both Bob and Charlie. Both of them can then use this information to perform the transformation $\mathcal{M}$ on their quantum systems. Because each of their respective systems was entangled with Alice's state in the beginning, the transformation will produce the initial state $\ket{\psi}$ in both Bob's and Charlie's lab, effectively cloning it. Because we have shown above that cloning unknown quantum states is impossible, in turn, it must be impossible for Alice's qubit to be maximally entangled with one system and at the same time maximally entangled with another system.
\section{Quantum key distribution}
\label{sec:QKDprotocols}

Quantum key distribution protocols enable two spatially distant parties to establish a shared secret key, which means that both parties have an identical, completely random bit string, such that the knowledge an adversary might have on the key is negligible. Combining such a protocol with the one-time pad scheme (see \cref{fig:onetimepad}) enables unconditionally secure data encryption. There is a variety of QKD protocols tailored to different aspects, but their security can always be traced back to the quantum phenomena that were explained in the preceding section. Some protocols are designed to account for complications in practical implementations such as imperfect devices and noisy quantum channels, and different kinds of protocols make different kinds of assumptions on the incorporated physical devices, resulting in varying levels of security. Here, we explain the general structure of a QKD protocol and present different classes of such protocols. Additionally, we discuss the state of the art in both theory and experiment, and which challenges have yet to be overcome.

\subsection{The structure of QKD protocols}
\label{subsec:protocol}

As a starting point, we describe a general QKD protocol that exploits entanglement between the two honest parties, Alice and Bob. Most protocols are varieties of this one, although it must be noted that the first proposed QKD protocol, the BB84 protocol \cite{Bennett1984}, had a different structure. Instead of using a source that distributes entangled quantum states to Alice and Bob, Alice prepares the quantum states (according to a random bit) in her laboratory and sends them to Bob, who subsequently measures them. This is known as a \emph{prepare and measure} (PM) scheme. Since one can formulate an equivalent entanglement-based (EB) protocol and reduce the security of the PM scheme to the security of the EB protocol, we focus here only on the latter type for simplicity.

\begin{figure}[t]
	\centering
	\begin{tikzpicture}[scale=1]
		\node at (-0.5,1.75) {\color{myred} \textsf{Alice}};
		\draw[fill=myred!40!white,blur shadow={shadow blur steps=10}] (-1,-1) rectangle (1.75,1.5);
		\draw[fill=LightGray,draw=black] (0.6,0.6) rectangle (1.5,1.25);
		\node at (1.05,0.925) {\small $\mathcal{Q}$};
		\draw[fill=LightGray,draw=black] (0.6,-0.75) rectangle (1.5,-0.1);
		\node at (1.05,-0.425) {\small $\mathcal{C}$};
		\draw[->,>=stealth] (1.05,0.5) -- (1.05,0);
		\draw[fill=LightGray,draw=black] (-0.75,-0.1) rectangle (0.15,0.55);
		\node at (-0.3,0.225) {\small $\mathtt{RNG}$};
		\draw[->,>=stealth] (-0.3,0.65) to[bend left] (0.5,0.925);
		\draw[->,>=stealth] (-0.3,-0.2) to[bend right] (0.5,-0.425);
		\draw[->,>=stealth] (1.05,-0.85) -- (1.05,-1.5);
		\node at (1.05,-1.75) {\small $S_A$};
		\begin{scope}[xshift=9cm,xscale=-1]
			\node at (-0.55,1.75) {\color{myblue} \textsf{Bob}};
			\draw[fill=myblue!40!white,blur shadow={shadow blur steps=10}] (-1,-1) rectangle (1.75,1.5);
			\draw[fill=LightGray,draw=black] (0.6,0.6) rectangle (1.5,1.25);
			\node at (1.05,0.925) {\small $\mathcal{Q}$};
			\draw[fill=LightGray,draw=black] (0.6,-0.75) rectangle (1.5,-0.1);
			\node at (1.05,-0.425) {\small $\mathcal{C}$};
			\draw[->,>=stealth] (1.05,0.5) -- (1.05,0);
			\draw[fill=LightGray,draw=black] (-0.75,-0.1) rectangle (0.15,0.55);
			\node at (-0.3,0.225) {\small $\mathtt{RNG}$};
			\draw[->,>=stealth] (-0.3,0.65) to[bend left] (0.5,0.925);
			\draw[->,>=stealth] (-0.3,-0.2) to[bend right] (0.5,-0.425);
			\draw[->,>=stealth] (1.05,-0.85) -- (1.05,-1.5);
			\node at (1.05,-1.75) {\small $S_B$};
		\end{scope}
		\begin{scope}[yshift=0.2cm]
			\draw[fill=LightGray,blur shadow={shadow blur steps=10}] (4.5,1.75) circle(0.75cm);
			\node at (4.5,1.85) {\footnotesize quantum};
			\node at (4.5,1.5) {\footnotesize source};
			\draw[decoration={snake},decorate,postaction={decoration={markings,mark=at position 1 with {\arrow[scale=1.5]{>}}},decorate}] (3.65,1.75) -- (1.85,0.75);
			\draw[decoration={snake},decorate,postaction={decoration={markings,mark=at position 1 with {\arrow[scale=1.5]{>}}},decorate}] (5.35,1.75) -- (7.15,0.75);
			\node at (2.7,1.5) {\rotatebox{30}{\footnotesize quantum}};
			\node at (2.9,1.1) {\rotatebox{30}{\footnotesize channel}};
			\begin{scope}[xshift=3.4cm]
				\node at (2.9,1.5) {\rotatebox{-30}{\footnotesize quantum}};
				\node at (2.7,1.1) {\rotatebox{-30}{\footnotesize channel}};
			\end{scope}
		\end{scope}
		\begin{scope}[yshift=-0.75cm]
			\draw[fill=LightGray,blur shadow={shadow blur steps=10}] (1.85,0.3) -- (2.1,0.5) -- (2.1,0.4) -- (6.9,0.4) -- (6.9,0.5) -- (7.15,0.3) -- (6.9,0.1) -- (6.9,0.2) -- (2.1,0.2) -- (2.1,0.1) -- cycle;
			\node at (4.5,0.65) {\small authenticated classical channel};
		\end{scope}
	\end{tikzpicture}
	\caption{\label{fig:QKDprotocol}\textbf{Setting of a typical QKD protocol.} A source distributes quantum systems in an entangled state to Alice and Bob via insecure quantum channels. The two parties each have access to a quantum device $\mathcal{Q}$, a classical computer $\mathcal{C}$, and a trusted random number generator $\mathtt{RNG}$. Additionally, Alice and Bob can communicate via an authenticated public classical channel. At the end of the protocol, the devices will output a key pair $(S_A,S_B)$ or abort.}
\end{figure}
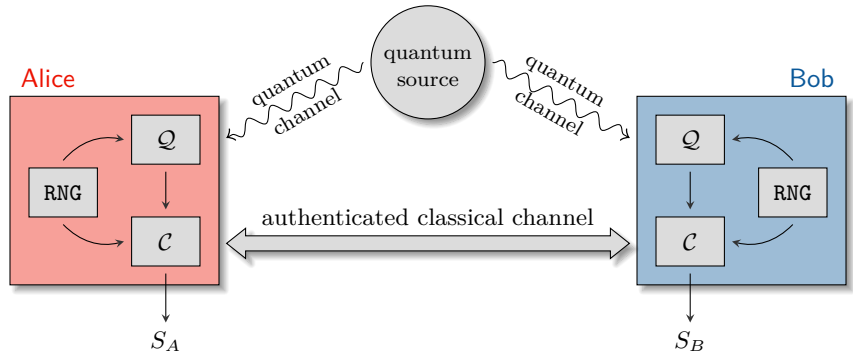

The goal of a QKD protocol is to either output a shared secret key which fulfills the desired properties, or to abort. It will never output an insecure key. The general setting as depicted in \cref{fig:QKDprotocol} is the following: Alice and Bob have access to a source (which can be located either inside one of Alice and Bob's laboratories or outside of both) that distributes quantum systems via an insecure quantum channel. This allows Eve to arbitrarily interact with the quantum systems. However, due to the nature of quantum mechanics she will either be detected, or not gain any information on the quantum states as explained in the preceding section. Furthermore, each party has a quantum device $\mathcal{Q}$ to measure the received quantum systems, a classical computing device $\mathcal{C}$ that stores and processes all classical information, and a trusted random number generator $\mathtt{RNG}$ that provides the randomness required for some steps of the protocol (see below). In addition, Alice and Bob can communicate via an authenticated public classical channel, which means that an eavesdropper can listen to all of their messages, but cannot modify them.

We divide the protocol into two phases: The first one is the \emph{quantum phase}, where quantum systems are distributed and measured, thus acquiring the classical data called the \emph{raw key}. The following two steps are repeated many times:
\begin{enumerate}
	\item \textbf{State distribution:} The source distributes quantum systems in an entangled state (e.g., the qubit state $\ket{\Phi}_{AB}$ from \cref{eq:Bell}) to Alice and Bob via insecure quantum channels. If the source is located in one of the laboratories, say Alice's lab, in this step Alice prepares the system in an entangled state and sends one half of it to Bob via an insecure quantum channel while keeping the other half. During this step, it is possible for an eavesdropper to tamper with the quantum systems.
	\item \label{item:bases}\textbf{Measurement:} Alice and Bob each randomly and independently of each other select one out of several possible measurements. The necessary randomness for this step is provided by their respective random number generators $\mathtt{RNG}$. For example, they could choose between measuring in the $\{\ket{0},\ket{1}\}$-basis and in the $\{\ket{+},\ket{-}\}$-basis (see \cref{subsec:statespace}). Both parties measure their part of the system using the quantum device $\mathcal{Q}$ and store the classical result. It is crucial that these choices are made randomly to avoid that the eavesdropper can predict the measurements and adapt her strategy accordingly.
\end{enumerate}
After completing the quantum phase, Alice and Bob share a pair of bit strings, which are partially correlated and partially secret. This stems from the fact that an eavesdropper could have performed an attack on the quantum systems during the protocol, thus compromising the entanglement and collecting information on the raw key. In addition, Alice and Bob have to choose their measurements randomly and independently of each other in the second step. Consequently, in those rounds where Alice and Bob have chosen incompatible measurements (for example, if Alice has chosen to measure in the $\{\ket{0},\ket{1}\}$-basis and Bob in the $\{\ket{+},\ket{-}\}$-basis), their results are uncorrelated.

All these aspects are accounted for in the second phase, the \emph{classical post-processing}. This part of the protocol is purely classical and it has two objectives: The first one is to estimate the errors in Alice and Bob's bit strings. Errors can occur because of noisy channels and imperfect devices, but also because of the attack of an eavesdropper as noted above. To account for the worst-case scenario, all errors are typically attributed to the involvement of an eavesdropper. The amount of errors decides whether Alice and Bob abort the protocol or continue with the classical post-processing. The second objective of this phase is to turn the generated raw key pair that Alice and Bob share into a pair of shorter but identical and secret bit strings, the \emph{cryptographic key}. The steps are the following:
\begin{enumerate}
	\setcounter{enumi}{2}
	\item \textbf{Sifting:} Alice and Bob publicly compare the bases they have chosen in Step~\ref{item:bases}. Only those rounds where they have chosen the same measurement provide correlated results, hence they discard those rounds where they have chosen different measurements.
	\item \textbf{Parameter estimation:} Alice and Bob randomly choose a number of rounds and announce their respective results. In an ideal scenario with perfect devices and no eavesdropper, the results should always coincide after the sifting step. Hence, the comparison of their results gives them an accurate estimate of the error rate of their strings, from which they can derive a bound on the information that an eavesdropper could have gained during the execution of the protocol. If the error rate is higher than a previously determined threshold, Alice and Bob abort the protocol.
	\item \textbf{Error correction:} After having identified an estimate on the error rate of their strings, Alice and Bob use a classical error correction protocol to turn their correlated strings into identical strings. Afterwards, they check whether the error correction procedure has been successful by comparing hashes of their respective bit strings (this is explained in more detail in \cref{subsec:security}).
	\item \textbf{Privacy amplification:} In the last step, Alice and Bob employ a privacy amplification procedure to remove Eve's information on the key, which produces a final key of shorter length. It is crucial that they have bounded Eve's knowledge accurately in the parameter estimation step, otherwise she will still have some knowledge on the final key. 
\end{enumerate}
After completing the steps of this protocol, one of two possible scenarios will occur: The protocol has aborted because Eve obtained too much information on the key. In this case, Alice and Bob simply restart the protocol and make a second attempt at generating a key. The second scenario is that they now hold a pair $(S_A,S_B)$ of identical, secret bit strings, which they can then use to encrypt messages. The protocol will never (or, to be precise, with negligible probability) output an insecure key.

\subsection{Quantifying security}
\label{subsec:security}

In contrast to classical key generation schemes, the structure of a QKD protocol allows us to quantify security. More precisely, it allows us to compute the security parameter $\epsilon^{\mathrm{QKD}}$ that was introduced in \cref{sec:Classicalcrypto}. It is important to note that the security is never perfect. For example, in the quantum phase, Eve could guess the measurement bases correctly for each protocol round. In this case, she could measure all the systems without being detected, and her outcomes would be perfectly correlated with Alice's information. Even though this is possible, it is highly unlikely. If the protocol has $n$ rounds, the probability of Eve guessing all bases correctly is vanishingly small, namely $2^{-n}$. As mentioned above, it is also possible that the protocol aborts because the eavesdropper has gained too much information on the raw key. In this case, Alice and Bob can simply restart the protocol and make a second attempt at generating a key. The scenario they want to avoid is that the protocol \emph{does not abort} but outputs an insecure key. An insecure key is a key where either the two bit strings of the key pair are not identical or where Eve has received information about the key. This results in two quantities that describe the security of a QKD protocol:
\begin{enumerate}
	\item $\epsilon_\mathrm{correct}$ is the probability that the protocol does not abort, but Alice and Bob's bit strings are not identical.
	\item $\epsilon_\mathrm{secret}$ is the probability that the protocol does not abort, but Eve has some information about the generated key pair.
\end{enumerate}

The values $\epsilon_\mathrm{correct}$ and $\epsilon_\mathrm{secret}$ are often combined to a single security parameter $\epsilon^{\mathrm{QKD}}\coloneqq\epsilon_\mathrm{correct}+\epsilon_\mathrm{secure}$. They are directly connected to the secret key rate of the protocol, which is the number of key bits generated per round. In the error correction step, Alice and Bob not only apply an error correction protocol, but also check whether it has been successful. The latter is typically done by comparing the outputs of a \emph{hash function} $f$ of the respective bit strings. These functions have two properties: (i) From the output of the function it is impossible to know what the input to the function was, and (ii) if the outputs $f(x),f(y)$ of two inputs $x$ and $y$ are equal, then (with high probability) the inputs $x$ and $y$ are also equal. Alice and Bob hence each apply the hash function $f$ (which is randomly chosen but publicly known) to their respective bit strings after the error correction procedure and publicly compare the outputs. If these are equal, with high probability the input strings are also equal. This procedure allows Alice and Bob to check that the error correction procedure has been successful without revealing information about their bit strings. The disadvantage is that there is a small but nonzero probability that the bit strings are different even though the corresponding hashes are identical. This probability depends on how many bits have been consumed during the checking procedure, resulting in a trade-off between the amount of key that is generated and the probability that the bit strings still contain errors. The smaller that probability is made, the more bits from the raw key must be used for error correction, hence shortening the final key. A similar situation presents itself in the privacy amplification step, where the error-corrected bit strings are compressed to remove Eve's information. The more they are compressed, the less information Eve has on them; hence it is a trade-off between the length of the key and its secrecy. Therefore, choosing suitable values for $\epsilon_\mathrm{correct}$ and $\epsilon_\mathrm{secret}$ means finding a trade-off between the amount of key generated and the level of security it guarantees. 

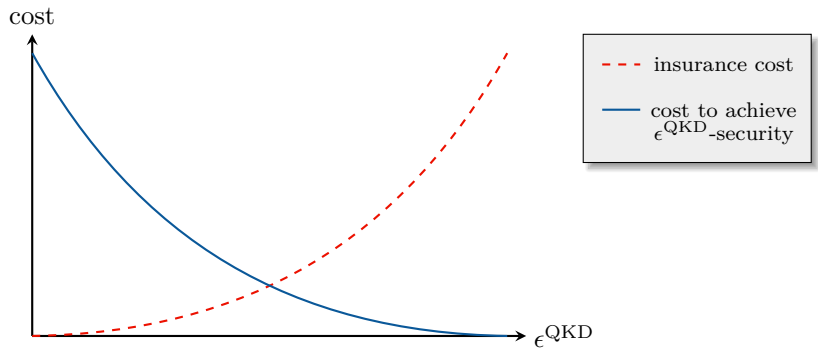
\begin{figure}[t]
	\centering
	\begin{tikzpicture}
		\draw[thick,->,>=stealth] (-0.5,0) -- (6,0);
		\draw[thick,->,>=stealth] (-0.5,0) -- (-0.5,4);
		\node at (6.5,0) {$\epsilon^\mathrm{QKD}$};
		\node at (-0.5,4.25) {cost};
		\draw[thick, color=myred, dashed] (-0.5,0) to [bend right] (5.75,3.75);
		\draw[thick, color=myblue] (-0.5,3.75) to [bend right] (5.75,0);
		\draw[fill=LightGray!50,blur shadow={shadow blur steps=10}] (6.75,2.3) rectangle (9.75,4);
		\draw[thick, color=myred, dashed] (7,3.6) -- (7.5,3.6);
		\node at (8.6,3.6) {\footnotesize insurance cost};
		\node at (8.6,3) {\footnotesize cost to achieve};
		\node at (8.6,2.7) {\footnotesize $\epsilon^\mathrm{QKD}$-security};
		\draw[thick, color=myblue] (7,3) -- (7.5,3);
	\end{tikzpicture}
	\caption{\label{fig:costvssecurity}\textbf{Schematic depiction of the trade-off between security and costs.} The graph shows the two types of costs associated to a given security level~$\epsilon^\mathrm{QKD}$: The cost of implementing a protocol that guarantees the corresponding level of security (depicted in blue, solid), and the cost for an insurance that would pay for the damage incurred by a security breach (depicted in red, dashed). Increasing the security level, which corresponds to lowering~$\epsilon^\mathrm{QKD}$, requires higher implementation costs, but lowers the cost for insurance (since a security breach is less likely). For example, $\epsilon^\mathrm{QKD}$ can be made smaller at the cost of increasing the length of the raw key, which in turn requires a higher communication bandwidth. An optimal choice of the parameter $\epsilon^\mathrm{QKD}$ minimizes the total of the implementation cost and the insurance cost.}
\end{figure}

What is a suitable value of the security parameter $\epsilon^\mathrm{QKD}$? As it represents the probability that something goes wrong without the protocol aborting, ideally we want it to be as close to zero as possible. To give some guidance for determining an appropriate value of the security parameter we can associate two kinds of costs to a particular value of $\epsilon^\mathrm{QKD}$ (see \cref{fig:costvssecurity}): One is the cost of implementing a protocol that guarantees the corresponding level of security, and the other it the cost of insuring possible malfunctions of the device. Decreasing the value of $\epsilon^\mathrm{QKD}$ is always accompanied by an increase in cost for the implementation of the protocol, because we need more rounds for the same amount of key when $\epsilon^\mathrm{QKD}$ decreases, but will give rise to a decrease of the insurance cost, since the probability of a security breach is smaller. Choosing the optimal value of $\epsilon^\mathrm{QKD}$ hence corresponds to minimizing the total of these two costs. Typical values for $\epsilon^\mathrm{QKD}$ per generated key range from $\epsilon=10^{-6}$ to $\epsilon=10^{-12}$, see for example \cite{ArnonFriedman2018, Tomamichel2017}.

\subsection{Assumptions and security guarantees}
\label{subsec:assumptions}

Any security proof we formulate for a QKD protocol is based on assumptions about the systems, states, and measurements and even about the underlying physical theory. Most of these assumptions depend on the kind of protocol that is considered, but there are some fundamental ones independent of this choice:
\begin{enumerate}
	\item \textbf{Quantum theory is correct.} We assume that quantum theory makes accurate predictions about measurement outcomes. This is supported by the fact that countless experiments have verified the predictions of quantum mechanics. Nevertheless, the following remark is in order here: Although the assumption that quantum mechanics is correct is surely sufficient for quantum cryptography to be secure, it might be too strong for this purpose. The security of quantum cryptography does not rely on all aspects of quantum theory, but only on those we have introduced in \cref{sec:Quantumtheory}: From the structure of the state space and operations on this space, plus the requirement that superluminal communication is impossible alone, we were able to show that quantum states cannot be cloned and entanglement is monogamous. We did not make use of other elements of quantum theory such as the Schrödinger equation. Consequently, the security of quantum cryptography only requires to assume that those parts of quantum theory discussed in \cref{sec:Quantumtheory} are correct. What's more, even if these parts of quantum theory have to be modified, quantum cryptography can still be expected to be secure as long as the changes are minor. This is especially important with regard to inevitable adjustments of quantum theory that will have to be made in order to combine it with gravity to a unified theory of quantum gravity. As long as those parts discussed in \cref{sec:Quantumtheory} are still approximately true, the security of quantum key distribution is still provable in a theory of quantum gravity.
	\item \textbf{Free randomness exists.} We assume that measurement choices (e.g., basis choices) can be made randomly and independently of the measurement device. This is indicated in \cref{fig:QKDprotocol} by the random number generator \texttt{RNG} and the quantum device $\mathcal{Q}$ being depicted as two independent devices.
	\item \textbf{Quantum theory is complete.} It has been shown that the completeness of quantum theory follows from its correctness and the existence of free randomness \cite{Colbeck2011}. We nevertheless list this point here independently of the others to emphasize its importance. Completeness means that there is no extension of the theory that can provide improved predictions. We have stated above that QKD guarantees security against an adversary who can use any attack that is possible within the framework of quantum theory. Completeness guarantees that the adversary cannot obtain any more information on the generated key than what is predicted by quantum theory. 
	\item \textbf{Devices do not leak any unauthorized information.} We assume that the incorporated devices, such as the quantum measurements, the random number generator, and the classical computer, only leak information as described in the protocol. For example, we assume that the raw key stored in the classical computer is not leaked in any way to the outside. To justify that this assumption is fulfilled, it is typically required that the lab is shielded accordingly.
\end{enumerate}

Apart from these fundamental assumptions about the underlying physical theory, there is a variety of additional assumptions one can make about the implementation of the protocol. Generally, the security proof becomes easier the more assumptions we make as each piece of information we have on the devices reduces the possible attacks an eavesdropper can perform. For example, if we assume that the devices do not leak any unauthorized information, we do not have to factor in this possibility in the security proof. On the other hand, the security proof only holds if the assumptions are fulfilled. Consequently, any deviation of the implementation from the assumptions renders the protocol effectively insecure. Depending on what assumptions are made in th protocol, it falls in one of three categories: device-dependent, device-independent, or semi-device-independent QKD. We explain these categories below.

\subsubsection*{Device-dependent QKD}

The first proposed quantum key distribution protocols, the BB84 protocol \cite{Bennett1984} and the Ekert protocol \cite{Ekert1991}, exactly specified which states have to be prepared and which measurements are carried out on those states. In the subsequent decades, further protocols were presented that also relied on the exact specification of the deployed devices for their security proof, see for example \cite{Bennett1992,Scarani2004}. These kinds of protocols fall under the name \emph{device-dependent} QKD, because the security proof depends on the exact characteristics of the devices in Alice and Bob's labs. For instance, the QKD protocol described in \cref{subsec:protocol} can be implemented using Bell states \cref{eq:Bell} and measurements in the $\{\ket{0},\ket{1}\}$- and the $\{\ket{+},\ket{-}\}$-basis for both Alice and Bob. Based on this choice, it is possible to calculate the key rate depending on the chosen values of $\epsilon_\mathrm{correct}$ and $\epsilon_\mathrm{secure}$. This is an example of a device-dependent protocol. 

While knowing the exact characterization of the incorporated devices simplifies the security proof of the protocol, it has some drawbacks: The resulting security statement only holds if the quantum devices behave exactly as specified, i.e., the quantum source produces only the predefined states and the measurement devices only measure in the two given bases. Consequently, any deviation of the actual devices in the lab compromises the protocol's security. In practice, however, the devices never behave exactly as specified. Some examples how implementations deviate from their theoretical description are that quantum states cannot be prepared with arbitrary precision, and measurement devices have dead times and dark counts. A more extensive list of problems regarding practical implementations can be found in \cite{Scarani2014}. 

This opens security loopholes (called side-channel attacks) the eavesdropper can exploit to obtain information about the key without being detected. This is known as \emph{quantum hacking} and has been demonstrated experimentally numerous times. What are typical attacks that exploit such loopholes? If the theoretical description of the protocol uses single photons as information carriers, the adversary can exploit that in practice, no perfect single-photon sources exist. Sometimes, the source emits a pulse consisting of two or more photons, and these pulses reveal information about Alice's basis choice that the adversary can exploit via the \emph{photon number splitting attack} \cite{Bennett1992b,Brassard2000,Felix2001}. As a countermeasure, the \emph{decoy-state} protocol \cite{Hwang2003,Wang2005,Lo2005} was developed that allows to detect this kind of attack. While this attack exploits vulnerabilities of the sender, the detectors are also susceptible to side-channel attacks. Typical examples are the \emph{time-shift attack} \cite{Makarov2006,Qi2007,Zhao2008}, which exploits inaccuracies of the photon detectors, and the \emph{detector blinding attack} \cite{Makarov2009,Lydersen2010,Gerhardt2011}, where the adversary gains control over the detectors by illuminating them with bright laser light. 

This mismatch between theoretical assumptions and realistic implementations has led to the development of QKD protocols that treat the quantum devices (or some of them) as black boxes instead of specifying which measurements they perform or which quantum states they produce. They fall under the name \emph{device-independent QKD} and \emph{semi-device-independent QKD}, and their security is based (either partly or solely) on phenomena that do not depend on any assumptions about the inner workings of the devices. Here, it is important to stress that this only concerns the quantum devices we employ in the protocol. The classical devices required for classical post-processing and storage of classical information as well as (to a certain extent) the random number generator always have to be trusted.
However, the higher security guarantees come at the cost that these protocols are technologically more demanding and hence harder to implement, which results in smaller achievable key rates and distances, as explained in \cref{subsec:stateoftheart}. In general, one has to find a trade-off between a reasonable set of assumptions regarding the resulting security and what is practically feasible.

\subsubsection*{Device-independent QKD}

\begin{figure}[t]
	\centering
	\begin{tikzpicture}[scale=1]
		\node at (-1.5,-2) {\textbf{\textsf{a)}}};
		\node at (-0.7,1.75) {\color{black} \textsf{Alice}};
		\draw[fill=LightGray!50!white,blur shadow={shadow blur steps=10}] (-1.2,-1) rectangle (1.75,1.5);
		\draw[fill=LightGray!90!black,draw=black] (-0.85,0.5) rectangle (0.05,1.15);
		\node at (-0.4,0.825) {\small $\mathtt{RNG}$};
		\draw[fill=LightGray!90!black,draw=black] (-0.85,-0.65) rectangle (0.05,0);
		\node at (-0.4,-0.325) {\small $\mathcal{C}$};
		\draw[->,>=stealth] (-0.4,0.4) -- (-0.4,0.1);
		\draw[fill=myred!40] (0.4,-0.25) rectangle (1.4,0.75);
		\draw[fill=myred!70] (0.65,0) rectangle (1.15,0.35);
		\node at (0.9,0.175) {\small \color{white} $a$};
		\begin{scope}[xscale=-1,xshift=-1.8cm]
			\draw[fill=white,draw=myred,densely dotted] (0.935,0.75) -- (1,0.9) -- (0.95,0.925) -- (0.875,0.75);
			\draw[fill=white,draw=myred,densely dotted] (1.01,0.975) circle (0.075cm);	
		\end{scope}
		\draw[fill=myred,draw=myred] (0.935,0.75) -- (1,0.9) -- (0.95,0.925) -- (0.875,0.75);
		\draw[fill=myred,draw=myred] (1.01,0.975) circle (0.075cm);
		\node at (0.9,1.25) {\small $x$};
		\draw[->,>=stealth] (-0.4,-0.75) -- (-0.4,-1.5);
		\node at (-0.4,-1.75) {\small $S_A$};
		\draw[->,>=stealth] (0.15,1) to[bend left] (0.7,1.25);
		\draw[->,>=stealth] (0.9,-0.35) to[bend left] node[below] {\small $a$} (0.15,-0.5);
		\begin{scope}[xshift=9cm,xscale=-1]
			\node at (-0.75,1.75) {\color{black} \textsf{Bob}};
			\draw[fill=LightGray!50!white,blur shadow={shadow blur steps=10}] (-1.2,-1) rectangle (1.75,1.5);
			\draw[fill=LightGray!90!black,draw=black] (-0.85,0.5) rectangle (0.05,1.15);
			\node at (-0.4,0.825) {\small $\mathtt{RNG}$};
			\draw[fill=LightGray!90!black,draw=black] (-0.85,-0.65) rectangle (0.05,0);
			\node at (-0.4,-0.325) {\small $\mathcal{C}$};
			\draw[->,>=stealth] (-0.4,0.4) -- (-0.4,0.1);
			\draw[fill=myred!40] (0.4,-0.25) rectangle (1.4,0.75);
			\draw[fill=myred!70] (0.65,0) rectangle (1.15,0.35);
			\node at (0.9,0.175) {\small \color{white} $b$};
			\begin{scope}[xscale=-1,xshift=-1.8cm]
				\draw[fill=white,draw=myred,densely dotted] (0.935,0.75) -- (1,0.9) -- (0.95,0.925) -- (0.875,0.75);
				\draw[fill=white,draw=myred,densely dotted] (1.01,0.975) circle (0.075cm);	
			\end{scope}
			\draw[fill=myred,draw=myred] (0.935,0.75) -- (1,0.9) -- (0.95,0.925) -- (0.875,0.75);
			\draw[fill=myred,draw=myred] (1.01,0.975) circle (0.075cm);
			\node at (0.9,1.25) {\small $y$};
			\draw[->,>=stealth] (-0.4,-0.75) -- (-0.4,-1.5);
			\node at (-0.4,-1.75) {\small $S_B$};
			\draw[->,>=stealth] (0.15,1) to[bend left] (0.7,1.25);
			\draw[->,>=stealth] (0.9,-0.35) to[bend left] node[below] {\small $b$} (0.15,-0.5);
		\end{scope}
		\begin{scope}[yshift=0.2cm]
			\draw[fill=myred!40,blur shadow={shadow blur steps=10}] (4.5,1.75) circle(0.75cm);
			\node at (4.5,1.85) {\footnotesize quantum};
			\node at (4.5,1.5) {\footnotesize source};
			\draw[draw=myred,decoration={snake},decorate,postaction={decoration={markings,mark=at position 1 with {\arrow[scale=1.5]{>}}},decorate}] (3.65,1.75) -- (1.85,0.75);
			\draw[draw=myred,decoration={snake},decorate,postaction={decoration={markings,mark=at position 1 with {\arrow[scale=1.5]{>}}},decorate}] (5.35,1.75) -- (7.15,0.75);
			\node at (2.7,1.5) {\rotatebox{30}{\color{myred} \footnotesize quantum}};
			\node at (2.9,1.1) {\rotatebox{30}{\color{myred} \footnotesize channel}};
			\begin{scope}[xshift=3.4cm]
				\node at (2.9,1.5) {\rotatebox{-30}{\color{myred} \footnotesize quantum}};
				\node at (2.7,1.1) {\rotatebox{-30}{\color{myred} \footnotesize channel}};
			\end{scope}
		\end{scope}
		\begin{scope}[yshift=-0.75cm]
			\draw[fill=LightGray,blur shadow={shadow blur steps=10}] (1.85,0.3) -- (2.1,0.5) -- (2.1,0.4) -- (6.9,0.4) -- (6.9,0.5) -- (7.15,0.3) -- (6.9,0.1) -- (6.9,0.2) -- (2.1,0.2) -- (2.1,0.1) -- cycle;
			\node at (4.5,0.65) {\small authenticated classical channel};
		\end{scope}
	\end{tikzpicture}\\\vspace{10pt}
	\begin{tikzpicture}[scale=1.05]
		\node at (-5.9,-1.25) {\textbf{\textsf{b)}}};
		\draw[fill=myred!40,blur shadow={shadow blur steps=10}] (-0.75,1.35) rectangle (0.75,2.35);
		\node at (0,1.85) {$\mathcal{B}$}; 
		\draw[->,>=stealth] (0,1.25) -- (0,0.85);
		\node at (0,0.7) {\small $m$};
		\node at (-1.6,1.7) {\rotatebox{30}{\color{myred} \footnotesize quantum}};
		\node at (-1.4,1.3) {\rotatebox{30}{\color{myred} \footnotesize channel}};
		\begin{scope}[xshift=3cm]
			\node at (-1.4,1.7) {\rotatebox{-30}{\color{myred} \footnotesize quantum}};
			\node at (-1.6,1.3) {\rotatebox{-30}{\color{myred} \footnotesize channel}};
		\end{scope}
		\begin{scope}[xshift=-4.1cm,scale=0.91]
			\draw[fill=LightGray,blur shadow={shadow blur steps=10}] (1.85,0.3) -- (2.1,0.5) -- (2.1,0.4) -- (6.9,0.4) -- (6.9,0.5) -- (7.15,0.3) -- (6.9,0.1) -- (6.9,0.2) -- (2.1,0.2) -- (2.1,0.1) -- cycle;
			\node at (4.5,-0.15) {\small authenticated classical channel};
		\end{scope}
		\node at (-4.75,2.25) {\color{black} \textsf{Alice}};
		\draw[fill=LightGray!50!white,blur shadow={shadow blur steps=10}] (-5.25,-0.25) rectangle (-2.5,2);
		\draw[fill=LightGray!90!black] (-3.25,0.875) circle (0.5cm);
		\node at (-3.25,0.875) {\footnotesize source};
		\draw[fill=LightGray!90!black] (-5,1.1) rectangle (-4.1,1.75);
		\node at (-4.55,1.425) {\small $\mathtt{RNG}$};
		\draw[fill=LightGray!90!black] (-5,0) rectangle (-4.1,0.65);
		\node at (-4.55,0.325) {\small $\mathcal{C}$};
		\draw[->,>=stealth] (-4.55,1) -- (-4.55,0.75);
		\draw[->,>=stealth] (-4,1.5) to[bend left] (-3.25,1.5);
		\draw[->,>=stealth] (-4.55,-0.1) -- (-4.55,-0.75);
		\node at (-4.55,-1) {\small $S_A$};
		\draw[draw=myred,decoration={snake},decorate,postaction={decoration={markings,mark=at position 1 with {\arrow[scale=1.5]{>}}},decorate}] (-2.4,1) -- (-0.85,1.85);
		\begin{scope}[xscale=-1]
			\node at (-4.8,2.25) {\color{black} \textsf{Bob}};
			\draw[fill=LightGray!50!white,blur shadow={shadow blur steps=10}] (-5.25,-0.25) rectangle (-2.5,2);
			\draw[fill=LightGray!90!black] (-3.25,0.875) circle (0.5cm);
			\node at (-3.25,0.875) {\footnotesize source};
			\draw[fill=LightGray!90!black] (-5,1.1) rectangle (-4.1,1.75);
			\node at (-4.55,1.425) {\small $\mathtt{RNG}$};
			\draw[fill=LightGray!90!black] (-5,0) rectangle (-4.1,0.65);
			\node at (-4.55,0.325) {\small $\mathcal{C}$};
			\draw[->,>=stealth] (-4.55,1) -- (-4.55,0.75);
			\draw[->,>=stealth] (-4,1.5) to[bend left] (-3.25,1.5);
			\draw[->,>=stealth] (-4.55,-0.1) -- (-4.55,-0.75);
			\node at (-4.55,-1) {\small $S_B$};
			\draw[draw=myred,decoration={snake},decorate,postaction={decoration={markings,mark=at position 1 with {\arrow[scale=1.5]{>}}},decorate}] (-2.4,1) -- (-0.85,1.85);
		\end{scope}
	\end{tikzpicture}
	\caption{\label{fig:M-DIQKD}Untrusted quantum devices. a)~In device-independent QKD, both the source of quantum states and the measurement devices, along with the quantum channels, are untrusted (indicated in red). The measurements are treated as black boxes with an input $x$ (respectively $y$ for Bob) and a classical output $a$ (respectively $b$). b)~In measurement-device-independent QKD, the preparation of quantum states takes place in Alice and Bob's respective laboratories. Both of them send a quantum system over an untrusted quantum channel to an untrusted relay where the measurement $\mathcal{B}$ takes place, whose outcome $m$ is publicly announced. Here, too, the untrusted parts are marked in red.}
\end{figure}
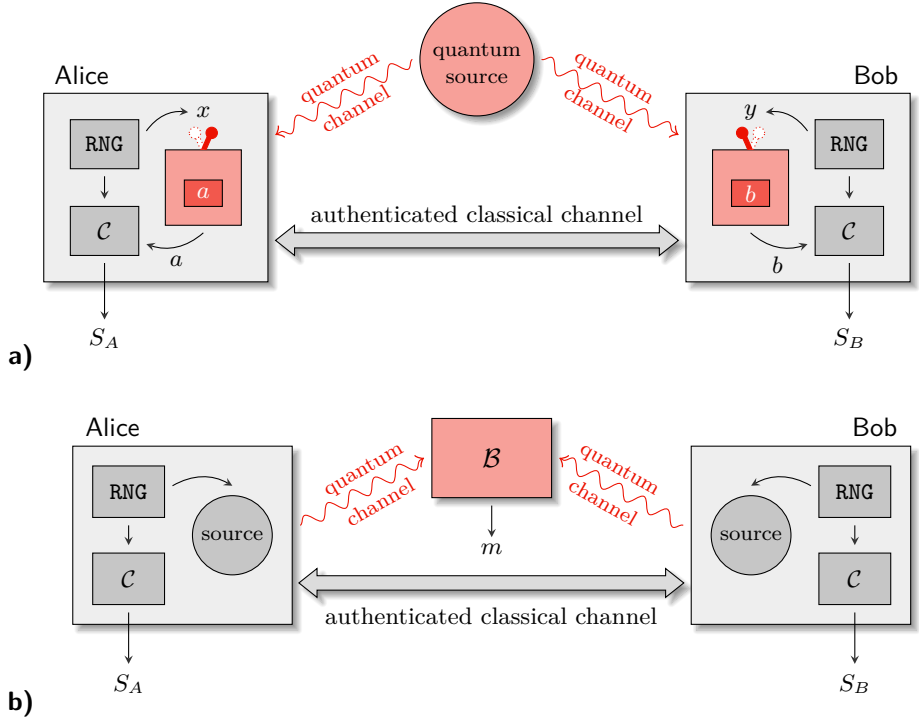

In device-independent QKD (DIQKD) protocols \cite{Pironio2009,Vazirani2014}, all quantum devices are treated as black boxes, see \cref{fig:M-DIQKD}a. First, this means that we do not make any assumptions about what quantum systems the source distributes to Alice and Bob. Second, we now view the measurement devices as boxes that take a classical input $x$ (respectively $y$ for Bob) and produce a classical output $a$ (respectively $b$). Although in practice, they are still realized by quantum measurements (where $x$ and $y$ correspond to the choice of measurement), the protocol's security is not derived from the precise states and measurements involved. Instead, the devices themselves are being tested throughout the execution of the protocol and the security is derived from these tests. This is typically achieved by evaluating a \emph{Bell inequality} \cite{Bell1964,Clauser1969}, which allows us to certify entanglement between Alice and Bob (and thus how much information an eavesdropper could have gained) simply from the input-output statistics of the boxes.

By removing all assumptions about the quantum devices from the security analysis, a higher level of safety is achieved, as the security proof applies to all possible implementations of the states and measurements involved in the protocol. As such, we do not have to worry about deviations from the theoretical description of the protocol. This even includes the case of malicious devices, where the manufacturer has manipulated the devices to obtain information about the key. Naturally, the improved security guarantees come at a cost. Not making any assumptions about the inner workings of the quantum devices also implies that we have to consider the worst case, where the eavesdropper, in principle, has absolute control over the devices. Consequently, we obtain lower key rates than in device-dependent settings, resulting in higher requirements on experimental implementations.

\subsubsection*{Semi-device-independent QKD}

As an alternative to treating all quantum devices as black boxes, it is possible to formulate protocols that are somewhere between device-dependent QKD (where all quantum devices are trusted) and DIQKD (where none of the quantum devices are trusted). A wide range of protocols can be imagined in this area: one can either trust the source that produces the quantum systems or the measurement devices, or trust only one of the involved parties. A variety of protocols have been formulated here, all of which fall into the category of \emph{semi-device-independent} QKD.

We cannot discuss all of these approaches in detail here, but one approach in particular is worth highlighting: Most attacks that exploit the implementation of a protocol aim at the detectors. To circumvent this problem, \emph{measurement-device-independent} (MDI) QKD \cite{Braunstein2012,Lo2012} rules out the possibility of exploiting imperfect detectors by removing them from the trusted part of the setting (i.e., Alice and Bob's respective labs) to an untrusted relay where the measurement takes place, see \cref{fig:M-DIQKD}b. Alice and Bob then independently prepare quantum states instead of receiving and measuring them. The measurement entangles the previously uncorrelated systems sent by Alice and Bob. The result of the measurement, together with the parties' respective local information about the prepared states, enables Alice and Bob to generate correlated key bits.

MDIQKD cannot provide the same level of security as DIQKD because the quantum state preparation has to be trusted. On the other hand, the characterization of the state preparation results in fewer requirements on experimental implementations and easier security proofs.

\subsection{Discrete- vs.\ continuous-variable QKD}

All of the protocols discussed above fall in the category of \emph{discrete-variable QKD} (DVQKD) because the information is encoded in a discrete way. Typically, this is implemented by using single photons as information carriers, exploiting for example different polarization directions to realize the two states of a qubit. The key is then established via detection of individual photons. In practice, however, there are no perfect single-photon sources and detectors, which has consequences for the security analysis of the protocol: If one does not account for this deviation from perfect single photons in the security proof, it opens loopholes that an adversary can exploit to obtain information on the key. An example of such an attack is the photon number splitting attack, which exploits that sometimes more than one photon is emitted. Including these deviations, on the other hand, generally makes the theoretical analysis more difficult and, additionally, decreases the key rate of the protocol.

These challenges have led to the idea of encoding information in properties of light that are continuous, such as the quadrature components of the electromagnetic field, yielding continuous values as measurement results. These protocols go by the name \emph{continuous-variable QKD} (CVQKD) \cite{Weedbrook2012}. The great advantage of this kind of protocols is that their implementation only requires standard telecommunication technology that is also used in classical optical communication. On the theoretical side, CVQKD faces the same challenges as DVQKD: A complete security proof includes proving security against general attacks under reasonable assumptions and for meaningful security parameters. The fundamentally different way of encoding information (namely via continuous instead of discrete quantities) however, leads to new challenges regarding the security analysis of such protocols because most of the information-theoretic techniques are specific to DVQKD and cannot directly be extended to the CV case.

\subsection{State of the art and challenges in theory and experiments}
\label{subsec:stateoftheart}

In recent years, enormous progress has been made in all areas of quantum cryptography, both in theory and experiment, with the aim of realizing a trustworthy QKD system with the best possible security guarantee. However, although QKD is at a completely different level today than it was ten years ago, there are still many challenges to overcome before it can be considered ready for the market, both on the theoretical and experimental side. 

In general, there are three parameters that allow us to benchmark and compare different protocols and implementations:
\begin{enumerate}
	\item \textbf{Security:} The most vital aspect of any quantum cryptographic primitive is security, which can be divided into three aspects. First, we need to state the assumptions we make on the incorporated devices, which means that we need to decide between device-dependent, semi-device-independent, and device-independent QKD. The second aspect is the kind of attacks we consider in the security proof: While allowing for the most general attacks by the eavesdropper provides the best security, these attacks are more complicated to analyze than those where the eavesdropper only has limited power, and the necessary techniques are developed to different levels for different kinds of protocols. The third aspect of security are the values for the two security parameters discussed in \cref{subsec:security}. Because they essentially represent the probability that the protocol fails without Alice and Bob being aware of it, choosing sensible values for them is a central part of the analysis of any protocol.
	\item \textbf{Achievable distance:} Another important criterion to decide whether the protocol is useful for a particular application is the distance over which a key can be established. Depending on the protocol and the implementation, this is limited by different factors. One main limitation, especially with regard to optical implementations, is the photon loss in optical fibers.
	\item \textbf{Key rate:} The rate at which a key is generated is a significant performance measure for any QKD protocol. A protocol that can generate a key over long distances is only of practical relevance if the key rate is sufficiently high. The long-term goal is to close the gap between typically deployed classical key rates (around 100\,Gbit/s) and those achievable with quantum protocols. Limiting factors here generally are the efficiency and dead time of the operated detectors. For protocols that use a source of entanglement, the rate at which entangled states can be successfully generated is another crucial aspect that limits the key rate.
\end{enumerate}

It is generally impossible to have optimal values for all three factors listed above: Increasing the distance between the two parties in a QKD protocol results in additional losses, which reduces the key rate. Likewise, requiring small $\epsilon$-values comes with additional costs for error correction and privacy amplification, also reducing the key rate. Therefore, a crucial part of any practical realization of a QKD protocol is to find a reasonable trade-off between these three aspects, depending on the focus of the implementation. In the remainder of this section, we present the state-of-the-art theoretical and experimental developments for the different kinds of protocols presented in the preceding section, and the main challenges that still have to be overcome.

\subsubsection{Security proofs}

On the theoretical side, the ultimate goal for any QKD protocol is to prove its security against general attacks where we assume that the eavesdropper is only limited in her attacks by the laws of physics. Because proving the security of the different types of protocols presented in \cref{subsec:assumptions} is increasingly challenging the fewer assumption we make, naturally the current state of the respective proofs is different.

In device-dependent QKD we have the most complete security proofs (see, for example, \cite{Tomamichel2017}). Given that we accept the assumptions we make in this case (such as the specification of the quantum states and measurements), there exist techniques for proving the security of these protocols against general attacks. A typical strategy for such a proof is to first analyze the protocol for \emph{collective attacks}, i.e., where the adversary applies the same attack in every round, and the individual rounds are uncorrelated. For this case, powerful numerical tools have been developed \cite{Coles2016,Winick2018}. From this, security against general attacks can then be inferred via techniques based on the \emph{quantum de Finetti theorem} \cite{Renner2008,Christandl2009} or the \emph{Entropy Accumulation Theorem} (EAT) \cite{Dupuis2020,George2022,Metger2022}.

In device-independent QKD, one can use a similar strategy to prove security. First, one can prove security against collective attacks using numerical methods such as the one presented in~\cite{Brown2021}. This can then be lifted to a full security proof using the EAT \cite{ArnonFriedman2018}, which is applicable to both device-dependent and device-independent protocols. However, the biggest issue that is unsolved to this day is \emph{composability} of DIQKD. A protocol is said to be composably secure if it can be used as a subprotocol in any other routine without compromising security. This is a natural requirement if the generated key is supposed to be used in another cryptographic protocol such as the one-time-pad scheme. While composable security can be shown for device-dependent QKD \cite{Portmann2021}, it has not yet been rigorously proven for device-independent protocols. Even worse, there is evidence that DIQKD is \emph{not} composably secure when the same devices are being reused in the composition. It was pointed out in~\cite{Barrett2013} that a malicious device could store data that was generated during the first execution of a protocol, and upon a second run of the protocol leak the data from the first run. Because all the techniques used to prove security only deal with a single execution of the protocol, these kind of attacks are not accounted for. Note that this is not a problem in the device-dependent case: Here, we can make the assumption that the devices do not store any information after the execution of the protocol is finished. This kind of assumption cannot be made in the DIQKD case because it is a direct assumption on the incorporated quantum devices and therefore contradicts the idea of device-independence.

If information is encoded in continuous variables instead of discrete ones we face new challenges regarding the theoretical analysis. Many techniques that are used in the discrete-variable case cannot be directly extended to CVQKD, such as the aforementioned quantum de Finetti theorem and the Entropy Accumulation Theorem. Consequently, there is no general security proof yet, only proofs for a limited number of protocols that use specific states and measurements. An overview of the current state of security proofs in CVQKD can be found in \cite{Diamanti2015}.

\subsubsection{Experimental implementations}

Before we discuss achievements in the respective categories of protocols, we want to mention one challenge that all of them face: There is a fundamental limit to the distance over which a key can be generated due to losses in the quantum channel that connects the two parties \cite{Takeoka2014,Pirandola2017}. One approach to overcome the distance limit is the use of quantum repeaters \cite{Briegel1998,Sangouard2011,Azuma2022}. Unlike classical signals, the quantum signals used in QKD cannot be noiselessly amplified due to the no-cloning theorem (see \cref{subsec:nocloning}). However, using quantum memories makes it possible to establish a pair of entangled states over high distances. The idea is to divide the quantum channel into segments (which can be made sufficiently small) and generate entangled pairs between the intermediate stations. These entangled pairs are then used to carry out the quantum teleportation protocol described in \cref{subsec:monogamy}, effectively teleporting entanglement from one party to the other via the intermediate stations. Even though this approach, in principle, allows to overcome the distance limit, it is not yet experimentally realized. Despite the considerable progress that has been made on implementing quantum repeaters (see, for example, \cite{Liu2021,LagoRivera2021}), the main problem is still that state-of-the-art quantum memories cannot store quantum systems for the time span that is needed to establish the entangled state between the two distant parties. It is worth noting that once quantum computers become available, they can be used as quantum memories in repeaters.

Another possible solution to this problem is offered by incorporating low-Earth orbit satellites into the implementation. These act as intermediate stations between the two parties who aim to establish a secret key on the ground. Because free-space communication is less susceptible to loss, in particular at high altitudes, it is possible to achieve longer distances with this technology compared to optical fiber connections. The use of satellites in QKD protocols has already been successfully demonstrated in \cite{Liao2018}, and there are ongoing efforts to incorporate satellites into QKD networks on a global scale \cite{Bedington2017,Sidhu2021}. Elevating this technology to a commercial level requires innovations both on the hardware side (as demonstrated in \cite{Ecker2021}) and with regard to security proofs (see, for example, \cite{Lim2021}). Quantum key distribution in space also has other possible applications, see, for example, \cite{QKDinSpace2020}.

\subsubsection*{Device-dependent QKD}

Device-dependent QKD has the advantage that the security proof is simpler since the complete characterization of the devices can be used. Therefore, it is not surprising that the experiments that have achieved the highest key rates and distances fall into this category. In 2018, a secret key rate of $13.72$\,Mbit/s over a distance of $10$\,km with security parameter $\epsilon=10^{-10}$ was reported in~\cite{Yuan2018}. Although this experiment demonstrated that high key rates are achievable in principle, the distance is too short for many practical purposes. An implementation that focuses on large distances was carried out in~\cite{Boaron2018} in 2018, where a record distance of $421$\,km was achieved with $\epsilon=10^{-9}$. However, at this distance, the key rate was only $0.25$\,bit/s, which is too small for practical applications.

These experiments demonstrate how much the distance affects the achievable secret key rate. A possible way to overcome this complication is to use low-Earth orbit satellites as links. Compared with terrestrial channels, satellite-to-ground communication has much smaller losses, which makes it a promising candidate for long-distance quantum communication. With this approach, it is possible to establish a secure communication channel over a distance of $7,600$\,km (the distance between Xinglong, China and Graz, Austria) as shown in 2018 \cite{Liao2018}.

It is important to emphasize that in these implementations, we always have to trust the devices (which includes the satellites). Hence, we need to be aware of potential attacks that aim at exploiting inevitable deviations from the protocol by any device that is part of the practical implementation and possibly take countermeasures.

\subsubsection*{Device-independent QKD}

Device-independent QKD can naturally guarantee the highest level of security as it requires the fewest assumptions on the implementation. Although this level of security is desirable for practical purposes, it comes at a cost: The requirements for practical implementations are higher in the device-independent case, in the sense that they can tolerate fewer noise and losses in the devices. It is therefore not surprising that it has taken until 2021 for the first experimental implementations to successfully generate a secret key in the device-independent setting as reported in \cite{Nadlinger2021,Zhang2021,Liu2021a}.

The difficulty of realizing DIQKD experimentally is reflected in the achieved levels of security, distance, and secret key rates. Of the three experiments \cite{Nadlinger2021,Zhang2021,Liu2021a} mentioned above, only the implementation reported in~\cite{Nadlinger2021} was shown to be secure against the most general kind of attacks. Furthermore, the experiments have been carried out over distances between and $3.5$\,m and $700$\,m, with secret key rates between $8.7\cdot10^{-4}$\,bits/s and $466$\,bits/s. These numbers are far away from what is needed for practical use, especially because long distances are paired with small key rates.

Various challenges still have to be overcome to achieve a parameter regime relevant to applications: One of the main problems is the high photon loss in experimental equipment such as filters and optical fibers, which increases with the distance. As discussed above, quantum repeaters could help overcome this problem. Additionally, the rate at which entangled states are generated, which is an upper bound to the rate at which the key can be generated, is still too low for most applications.

\subsubsection*{Semi-device-independent QKD}

The fact that semi-device-independent QKD represents a compromise between security and practicality is reflected in the achievable distances and key rates. The current record for measurement-device-independent protocols, demonstrated in~\cite{Yin2016} in 2016 is a distance of $404$\,km with a key rate of $3.2\cdot10^{-4}$\,bit/s. Additionally, for smaller distances the group was able to obtain much higher key rates; for example, over a distance of $207$\,km a key rate of $9.55$\,bit/s was achieved. These numbers illustrate that the additional information we have on the devices in MDIQKD (compared to DIQKD) makes the experimental realization much easier, although not as easy as in the device-dependent case.

\section{Outlook}
\label{sec:Outlook}

Classical cryptographic protocols suffer from the fact that their security relies on the conjectured hardness of specific problems. Consequently, advancements in hardware and software development always constitute potential threats to their security. In contrast, quantum cryptography does not share the same fate. Taking advantage of the unique properties of quantum mechanics, security is based solely on the laws of physics. This means that the encryption cannot be broken regardless of how much power an adversary has.

One could argue that the disadvantage of quantum cryptography lies in the fact that it is expensive and difficult to realize. Although it is undoubtedly true that quantum cryptography poses many challenges on experimental implementations, it is important to note that it is technologically strictly simpler to realize than quantum computers. One of the most significant missing elements for long-distance experiments are reliable quantum memories, which are needed to build quantum repeaters that can amplify the quantum signals over large distances. These are also needed for quantum computers, but their requirements are much lower in QKD applications: Both the number of qubits and the time for which quantum states need to be stored are shorter in QKD than what is needed to build a universal quantum computer.

A recent publication by the NSA \cite{NSAwhitepaper} describes five main technical challenges that have to be overcome in order to make quantum key distribution market-ready. We want to briefly comment on these challenges and the prospect of QKD in meeting them.
\begin{enumerate}
	\item \emph{Critique: QKD is only a partial solution in the sense that it requires a way to authenticate the honest parties. The confidentiality offered by QKD can also be provided by post-quantum cryptography, which is typically less expensive with a better understood risk profile.}
	
	Initial authentication is a fundamental problem, which cannot be solved with any cryptographic method (classical or quantum) if the parties have no information about each other. Clearly, Alice cannot know that a message comes from Bob if there is nothing that distinguishes him from Eve.  Conversely, if Alice and Bob share a small initial secret (e.g., a password) then  authenticated communication between them can be achieved with existing classical cryptographic methods, even with information-theoretic security~\cite{Renner2004}. In a communication network, authenticated communication between two parties who share no initial information may be established, but requires a trusted third party. Independently of the method that is used for authentication (e.g., if it is only computationally secure),  QKD remains future-proof. The authentication has to be broken in real-time; Once the key is generated, breaking the authentication does not provide any information on the key.
	
	Regarding the second aspect of the critique, it is true that post-quantum cryptography is cheaper than QKD since it can easily be incorporated in current communication systems via software updates. However, the risk profile of post-quantum cryptography is only as well understood as the one of classical cryptography, which means that it relies on the presumed hardness of certain mathematical problems and thus always bears the risk of being broken by a classical or a quantum computer. This is demonstrated by the recent breach of one of the alternative finalists of the post-quantum cryptography challenge run by NIST \cite{NIST}. The risk profile of QKD, on the other hand, is understood perfectly since it comes with a mathematical proof as depicted in \cref{fig:quantcost}.
	
	\item \emph{Critique: QKD requires special purpose equipment and cannot be used with current hardware.} 
	
	This statement is correct with respect to the current state of communication hardware. However, the evolution of classical communication equipment will eventually come to the point where so much traffic has to be sent over the fibers that it is most efficient to encode each bit into a single photon, hence it will approach the kind of technology required by quantum communication. Anyway, considering that QKD is the only technology that can provide unbreakable secure encryption, it might be worth investing in more expensive hardware. This kind of technology will also not need any security patches since its security parameters are not influenced by developments in software or hardware. Compared to classical cryptography, which needs to be updated frequently to maintain its security standard (see \cref{fig:quantcost}), the QKD implementation is easier to maintain in this respect.
	
	\item \emph{Critique: QKD networks require trusted intermediate stations, which creates additional costs and security risks.} 
	
	This is true for current implementations of QKD over large distances but will be resolved with the development of quantum repeaters. These do not have to be trusted, but admittedly, they still cost.
	
	\item \emph{Critique: There is a difference between the security of the theoretical protocol and the security provided by the actual implementation.} 
	
	While this is true for device-dependent QKD protocols (and also classical cryptography, one should add), where all devices need to be characterized, it is not an issue anymore for device-independent QKD. Here, we do not have to make any assumptions about the quantum devices; hence, the security of the theoretical description does not differ from the one of the implementation. While this technology is still in a preliminary state, it provides a clear path towards full security of the implementations.
	
	\item \emph{Critique: QKD is more vulnerable to denial-of-service attacks.} 
	
	Here, one should not compare point-to-point QKD implementations to large classical communication networks. In a network, communication can be rerouted if one of the links fails to function. Since the ultimate goal is to build a QKD network with many nodes, this problem will eventually be resolved.
\end{enumerate}

The above discussion shows that there is a clear path to resolve all technical issues of QKD pointed out in \cite{NSAwhitepaper}. QKD can then offer a true advantage over classical cryptography. In computationally secure cryptography, the development of a new encryption algorithm is always followed by the search for an algorithm that breaks it, which itself is followed by the search for yet another secure encryption algorithm, forming a vicious circle. Quantum cryptography breaks this circle: It ensures that the encryption can never be broken regardless of any developments in hardware and software, even taking quantum computers into account. Hence, quantum protocols are not only secure during their execution, but information encrypted today will remain secure forever, independent of developments in (quantum) software and hardware. Accordingly, quantum cryptography can guarantee everlasting security.

\section*{Acknowledgements}

We thank Tristan Nguyen for his encouragement to write this article. We also thank Thomas Cope for helpful comments on the manuscript and Martin Bohmann for his input regarding satellite-based QKD. The work was supported by the Air Force Office of Scientific Research (AFOSR), grant No.~FA9550-19-1-0202, the Swiss National Science Foundation via the QuantERA project eDICT and via the National Centres of Competence in Research QSIT and SwissMAP, and the Quantum Center at ETH Zurich. 

\bibliographystyle{halpha}
\bibliography{QKDarticleBib}

\end{document}